\documentclass{SciPost}

\hypersetup{
    colorlinks,
    linkcolor={red!50!black},
    citecolor={blue!50!black},
    urlcolor={blue!80!black}
}

\usepackage[bitstream-charter]{mathdesign}

\usepackage{float,ulem}
\usepackage{graphicx,epsfig,epstopdf,amsmath}
\usepackage{xcolor}

\newcommand{\nn}{\nonumber\\}
\newcommand{\be}{\begin{equation}}
\newcommand{\ee}{\end{equation}}
\newcommand{\hT}{\hat{T}}
\newcommand{\hu}{\hat{u}}
\newcommand{\ct}{\Tilde{\chi}}
\newcommand{\tilt}{\Tilde{\theta}}
\newcommand{\dtemp}{\delta T}
\newcommand{\du}{\delta u}
\newcommand{\dhat}{\Hat{D}}
\newcommand{\ghat}{\Hat{\nabla}}

\DeclareSymbolFont{usualmathcal}{OMS}{cmsy}{m}{n}
\DeclareSymbolFontAlphabet{\mathcal}{usualmathcal}

\fancypagestyle{SPstyle}{
\fancyhf{}
\lhead{\colorbox{scipostblue}{\bf \color{white} ~SciPost Physics }}
\rhead{{\bf \color{scipostdeepblue} ~Submission }}

\fancyfoot[C]{\textbf{\thepage}}
}

\begin{document}

\pagestyle{SPstyle}

\begin{center}{\Large \textbf{\color{scipostdeepblue}{
Frame transformation and stable-causal hydrodynamic theory\\
}}}\end{center}

\begin{center}\textbf{
Sayantani Bhattacharyya\textsuperscript{1,2$\star$},
Sukanya Mitra\textsuperscript{2$\dagger$} and
Shuvayu Roy\textsuperscript{2$\ddagger$}
}\end{center}

\begin{center}
{\bf 1} School of Mathematics, University of Edinburgh, United Kingdom
\\
{\bf 2} School of Physical Sciences, National Institute of Science Education and Research, An OCC of Homi Bhabha National Institute, Jatni-752050, India.
\\[\baselineskip]
$\star$ \href{mailto:email1}{sayanta@niser.ac.in}\,,\quad
$\dagger$ \href{mailto:email1}{sukanya.mitra@niser.ac.in}\,,\quad
$\ddagger$ \href{mailto:email2}{\small shuvayu.roy@niser.ac.in}
\end{center}

\section*{\color{scipostdeepblue}{Abstract}}
\textbf{\boldmath{%
In this work, a connection has been indicated between the different existing formulations of relativistic hydrodynamic theories, which, in order to be causal and stable, (i) either requires `non-fluid' variables apart from velocity and temperature to be promoted to new degrees of freedom, or, (ii) needs to be in a generalized hydrodynamic frame other than those given by Landau or Eckart. The BDNK stress tensor (originally in a general frame) has been rewritten in the Landau frame using linearized all-order gradient-corrected redefinitions of the temperature and velocity fields. The redefinitions indicate that, while the BDNK formalism has a finite number of derivatives in the general frame, when written in the Landau frame, it either has an infinite number of derivatives, or one has to introduce MIS-like `non-fluid' variables by summing the infinite number of derivatives in the field redefinitions. There can be non-unique ways of performing these infinite-order summations. Finally, the dispersion relations and the corresponding spectra of these different systems of MIS type equations have been analyzed to check that the systems of equations presented here are indeed equivalent to the BDNK formalism, at least in the hydrodynamic regime.
}}

\vspace{\baselineskip}

\noindent\textcolor{white!90!black}{%
\fbox{\parbox{0.975\linewidth}{%
\textcolor{white!40!black}{\begin{tabular}{lr}%
  \begin{minipage}{0.6\textwidth}%
    {\small Copyright attribution to authors. \newline
    This work is a submission to SciPost Physics. \newline
    License information to appear upon publication. \newline
    Publication information to appear upon publication.}
  \end{minipage} & \begin{minipage}{0.4\textwidth}
    {\small Received Date \newline Accepted Date \newline Published Date}%
  \end{minipage}
\end{tabular}}
}}
}


\vspace{10pt}
\noindent\rule{\textwidth}{1pt}
\tableofcontents
\noindent\rule{\textwidth}{1pt}
\vspace{10pt}


\section{Introduction}
\subsection{A brief outline of relativistic dissipative hydrodynamics}
Relativistic fluid dynamics has been proven to be a reasonably successful theory describing the long wavelength collective behaviour of a range of physical systems - from cosmology and astrophysics \cite{Weinberg:1972kfs,Rezzolla:2013dea} to high-energy nuclear physics and ultrarelativistic heavy-ion applications \cite{Romatschke:2017ejr,Huovinen:2006jp,Heinz:2013th}. Conventional fluid dynamics is a classical effective field theory for large systems near thermodynamic equilibrium, where by a set of coupled, partial differential equations (PDEs), the state variables of the system describe the dynamical evolution of the conserved quantities at low energy, long wavelength limit \cite{Landau}.
It always comes with a cut-off (length) scale (determined by the underlying microscopic theory) above which the fluid-like descriptions are applicable. The variations (or derivatives with respect to the space and time) of the fluid variables measured in the units of the cut-off scale must be small and, therefore, can be treated perturbatively.
To have a schematic understanding, let us introduce the quantity called Knudsen number $K_n$ from the  kinetic theory description of a collective system of particles \cite{Degroot,Cercignani}. It is a dimensionless ratio of the particle mean free path $\lambda$ (which provides the microscopic length scale $l_{micro}$ of the system that is decided by the inter-particle interactions) and the macroscopic length scale ($L_{macro}$) over which the thermodynamic state variables considerably change,
\be
K_n\sim \lambda~\partial \sim l_{micro}/L_{macro}~,
\ee
where $\partial$ denotes derivative over the hydrodynamic fields which is inverse to the macroscopic length scale ($L_{macro}$). In an out of equilibrium scenario, the hydrodynamic fields (like energy-momentum stress tensor $T^{\mu\nu}$ or particle/charge current $J^{\mu}$) are expressed as an expansion in the powers of such a parameter that is required to be small (like $K_n<1$), and the resulting theoretical framework is formulated by a systematic build up of gradients of the fluid variables (such as temperature $T$, four-velocity $u^{\mu}$ and chemical potential $\mu$). The conservation laws of these hydro fields ($\partial_{\mu}T^{\mu\nu}=0$ and $\partial_{\mu}J^{\mu}=0$) provide the necessary equations of motion that describes the macroscopic evolution of the system under consideration. The microscopic information enters into the picture via the constants of proportionality of the field expansion equations called the transport coefficients.

Given the scenario, in principle, any realistic fluid description must have an infinite number of derivatives of the fluid variables in the equations of motion. However, such equations with infinitely many derivatives are of little practical use since it is impossible to solve them even numerically for any generic initial and boundary conditions. We need to truncate such infinite series, and most of the time, it turns out that arbitrary truncations lead to pathologies. For example, it is well known that if we truncate the relativistic fluid equations at the first subleading order in the derivative expansion with the most popular hydrodynamic frame choices prescribed by Landau and Eckart (the relativistic Navier-Stokes (N-S) equation that captures the leading effects of dissipation \cite{Landau,Eckart:1940te}), the equations admit solutions with superluminal signal propagation and thermodynamic instability \cite{Hiscock:1985zz,Hiscock:1987zz}.

In fact, recent analysis \cite{Heller:2022ejw,Hoult:2023clg} indicates that including non-hydrodynamic modes is probably essential to ensure the causality of the fluid equations (N-S lacks any \footnote{All of our analyses in this paper are performed in the local rest frame (the Lorentz frame where the background equilibrium fluid is in rest). All the comments made about the existence or non-existence of the non-hydrodynamic modes are made in this context unless explicitly mentioned otherwise.}). Here, by non-hydrodynamic modes, we refer to modes that have nonzero frequencies even in the absence of any spatial momenta. Such modes are named non-hydrodynamic for the following reasons. Hydrodynamics could also be viewed as the collective dynamics of the massless modes of a system, slightly away from the global equilibrium. Now, a large thermal system will be in  `global equilibrium' when the charge densities (or, more precisely, the corresponding chemical potentials) are constant over space so that the different infinitesimal subregions of the system are all in thermal equilibrium with each other. Near equilibrium dynamics (slow variation in time, relaxing towards global equilibrium) will be generated only when these charge densities or chemical potentials have variations in space. From this perspective, all fluid or hydro modes, encoding the near equilibrium dynamics, are the ones whose frequency or the variation in time vanishes as soon as the variation in space or the spatial momentum is set to zero. Any mode that does not satisfy this condition is a non-hydrodynamic mode.

Now, if causality necessarily requires the inclusion of non-hydrodynamic modes, one might think that in order to have a causal set of equations, one has to introduce non-hydrodynamic variables (variables that are not related to any conserved quantity). The well-known solution with this approach is the construction of Muller-Israel-Stewart (MIS) theory \cite{Muller,Israel:1976efz,Israel:1976tn,Israel:1979wp,Denicol:2012cn,Muronga:2001zk} where the shear tensor is introduced as a new variable (which is not associated with any thermodynamic conserved charge and has no equilibrium counterpart), with an extra equation of motion for it. The combined set of equations for the energy density $\varepsilon$ (as well temperature $T$), the fluid 4-velocity $u^{\mu}$ and the newly promoted variable shear tensor ($\pi^{\mu\nu}$) can have a finite number of derivatives and still predict causal propagation with stable perturbation decay \cite{Hiscock:1983zz,Olson:1990rzl,Pu:2009fj}. On the other hand, if we choose to integrate out the $\pi^{\mu\nu}$ by solving it first in terms of the fluid variables (using the perturbative technique of derivative expansion), the resultant fluid equations turn out to have an infinite number of derivatives (see \ref{section MIS}). Truncating this derivative correction at any finite order, however high it is, leads to a theory that is acausal, unless the corrections are infinitely summed up to all orders \cite{Mitra:2023ipl}.
 
Recently in \cite{Bemfica:2017wps,Bemfica:2019knx,Bemfica:2020zjp,Kovtun:2019hdm,Hoult:2020eho,Hoult:2021gnb,Rocha:2022ind,Biswas:2022cla,Biswas:2022hiv} another interesting way of constructing the causal-stable fluid theory (popularly known as the Bemfica-Disconzi-Noronha-Kovtun (BDNK) theory) has been introduced which has a finite number of derivatives but no extra `non-fluid' type variables. That means the BDNK hydro formalism does not require any additional degrees of freedom apart from the basic fluid variables $T$, $u^{\mu}$ and $\mu$ in order to preserve the causality and stability even in a finite order theory. The BDNK theory has achieved this by exploiting the freedom of field redefinition in effective field theories.
It is pathology-free only in a general frame (other than Landau or Eckart frame) and the out-of-equilibrium hydrodynamic variables are defined through their postulated constitutive relations that include both time and space gradients.
Here, the issue is discussed with further details below.

Velocity and temperature (or energy) are well defined in global equilibrium, but that is not the case once the system is away from it (see \cite{Dore:2021xqq,Kovtun:2012rj} for a detailed discussion on this point).
The derivative corrections to the equations of motion do not make sense unless we can precisely define what we mean by the fluid variables in derivative order. So, the standard strategy is to first define the fluid variables in terms of some microscopic quantity (field theory operators) and then explore the structure of the equations and their consequences.
In MIS theory, the velocity and energy density are defined via the `Landau frame' condition on the stress tensor $T^{\mu\nu}$ - the microscopic field theory operator. Here, the fluid velocity is the unique timelike unit-normalized eigenvector of the stress tensor, and the energy density is the corresponding eigenvalue. It has a nice physical interpretation: the velocity of the fluid is actually the velocity of the energy flow. Instead, if someone chooses the Eckart frame, the fluid velocity is defined in terms of the particle flow.

The BDNK approach deviates from this strategy of defining the fluid variables first. They explored the question of whether there exists any definition of the fluid variables such that the equations with a finite number of derivatives and also with no extra non-fluid variables could be causal. They then find a class of theories satisfying such conditions. So, in the BDNK formalism, we do have a tractable set of differential equations involving only the basic fluid variables like velocity and temperature, but we do not exactly know what this velocity or temperature means in terms of any measurable microscopic operator (like the stress tensor) once the system is away from the global equilibrium or ideal fluid limit.
\subsection{Summary and discussion of our results}
In this manuscript, our goal is to rewrite the BDNK stress tensor in the Landau frame by redefining the velocity and the energy density (temperature). In some sense, the key result of this work is the relation between the fluid variables in BDNK formalism (denoted by $u^\mu$ and $T$ respectively) and the velocity and the temperature field defined after frame transformation that are fixed through the Landau gauge condition (denoted as $\hat u^\mu$ and $\hat T$). We have explicitly worked out the relation for those fluid profiles that have small perturbations around some global equilibrium. We have assumed that the amplitudes of the perturbations are small enough so that a linearized treatment is justified. Further, in order to obtain an analytically tractable all order theory, we have restricted our analysis only to conformal, uncharged fluids in BDNK formalism.

To state our results in terms of equations, let us first introduce a notation $u^\mu -\hat u^\mu = \delta u^\mu$ and $T-\hat T= \delta T$. We have found that the shift variables $\delta u^\mu$ and $\delta T$ must satisfy the following differential equations up to terms that are linear in $\delta T$, $\delta u^\mu$ and their derivatives,
\begin{align}
&\frac{\delta T}{\hat{T}}+\tilde{\chi}\left[\frac{\hat{D}\hat{T}}{\hat{T}}
+ \frac{\hat{\nabla}_{\mu}\hat{u}^{\mu}}{3}\right]+\tilde{\chi}\left[\frac{\hat{D}\delta{T}}{\hat{T}}+\frac{\hat{\nabla}_{\mu}\delta{u}^{\mu}}{3}\right]=0~,
\nn
 &\delta u^{\mu}+\tilde{\theta}\left[\hat{D}\hat{u}^{\mu}+\frac{\hat{\nabla}^{\mu}\hat{T}}{\hat{T}}\right]+\tilde{\theta}\left[\hat{D}\delta{u}^{\mu}+\frac{\hat{\nabla}^{\mu}\delta{T}}{\hat{T}}\right]=0~.
 \label{shifteqn}
\end{align}
Next, we develop a formal solution for the equations \eqref{shifteqn} using two different methods. In both cases, it is manifested that the solutions will have terms up to all orders in derivative expansion. Finally, we introduce a set of new tensorial `non-fluid' variables (like the shear tensor in MIS theory) in order to recast the BDNK theory in an MIS-type formalism where the fluid variables like velocity and the temperature are defined through the Landau gauge condition.

In the first method, the equivalent system of equations will have an infinite number of `non-fluid' variables with the following nested structure of the energy-momentum tensor $T^{\mu\nu}$:
\begin{align}
 &\partial_\mu T^{\mu\nu}=0~,~~~~~~~~
 T^{\mu\nu}=\hat{\varepsilon}\left[\hat{u}^{\mu}\hat{u}^{\nu}+\frac{1}{3}\hat{\Delta}^{\mu\nu}\right]+\hat{\pi}^{\mu\nu},\nn
 &(1+\tilde\theta\hat D)\hat{\pi}^{\mu\nu}=-2\eta\hat{\sigma}^{\mu\nu}+\rho_1^{\mu\nu},\nn
 &(1+\tilde\chi\hat D)\rho^{\mu\nu}_1=(-2\eta)(-\tilde{\theta})\frac{1}{\hat{T}}\hat{\nabla}^{\langle\mu}\hat{\nabla}^{\rangle\nu} \hat{T}+\rho_2^{\mu\nu},\nn
 &(1+\tilde\theta\hat D)\rho^{\mu\nu}_2=(-2\eta)(-\tilde{\theta})\left(-\frac{\tilde{\chi}}{3}\right)\hat{\nabla}^{\langle\mu}\hat{\nabla}^{\rangle\nu} \hat{\nabla}\cdot\hat{u}+\rho_3^{\mu\nu},\nn
 &(1+\tilde\chi\hat D)\rho^{\mu\nu}_3=(-2\eta)(-\tilde{\theta})^2\left(-\frac{\tilde{\chi}}{3}\right)\frac{1}{\hat{T}}\hat{\nabla}^{\langle\mu}\hat{\nabla}^{\rangle\nu} \hat{\nabla}^2 \hat{T}+\rho_4^{\mu\nu},\nn
 &(1+\tilde\theta\hat D)\rho^{\mu\nu}_4=(-2\eta)(-\tilde{\theta})^2\left(-\frac{\tilde{\chi}}{3}\right)^2\hat{\nabla}^{\langle\mu}\hat{\nabla}^{\rangle\nu}\hat{\nabla}^2\hat{\nabla}\cdot\hat{u}+\cdots\nn
 &\vdots~~~~~~
 \label{MIS-type1intro}
\end{align}
In the second method, we need to introduce only one `shear tensor' type non-fluid variable, but its equation of motion turns out to be second order in spatial and third order in temporal derivatives,
 \begin{align}
 &\partial_\mu T^{\mu\nu}=0~,~~~~~~
 T^{\mu\nu}=\hat{\varepsilon}\left(\hat{u}^{\mu}\hat{u}^{\nu}+\frac{1}{3}\hat{\Delta}^{\mu\nu}\right)+\hat{\pi}^{\mu\nu}~,
 \nn
&\left[(1+\tilde{\theta}\hat{D})(1+\tilde{\chi}\hat{D})-\tilde{\theta}\frac{\tilde{\chi}}{3}~\hat{\nabla}^2\right]
\left\{(1+\tilde{\theta}\hat{D})\hat{\pi}^{\mu\nu}+2\eta\hat{\sigma}^{\mu\nu}\right\}
\nn
&~~~~~~~~~~~~~ =~
2\eta\tilde{\theta}\left\{1+(\tilde{\theta}+\tilde{\chi})\hat{D}\right\}\frac{\hat{\nabla}^{\langle\mu}\hat{\nabla}^{\nu\rangle}\hat{T}}{\hat{T}}~.
\label{MIS-type2}
\end{align}
We have analyzed the spectrum of linearized perturbations in both systems and found that all the hydrodynamic modes match those of the BDNK theory. This indicates that in the regime where fluid descriptions are applicable, all three systems of equations presented here are equivalent. However, equations \eqref{MIS-type1intro} and equations \eqref{MIS-type2} also have some extra non-hydrodynamic modes which are not there in the BDNK theory. The emergence of these new modes is possibly connected with the zero modes in the equations of the field redefinition (equations \eqref{shifteqn}) themselves.

Our equations are by no means more tractable than that of the BDNK. But here, the fluid variables have a clear and standard meaning, and since the velocity and temperature in BDNK theory could be precisely transformed to these variables (though we have derived it only at a linearized level), it attaches a similar definition to the BDNK fluid variables as well. Our analysis suggests that even in BDNK theory, there will be hidden non-fluid variables (or an infinite number of derivatives) if one would like to express the theory in terms of fluid variables only, which are locally defined through stress-energy tensor as we have in `Landau frame' \footnote{In this context, we should mention the analysis in \cite{Dore:2021xqq} where the authors connect the MIS and BDNK type formalisms with field redefinitions. However, while the authors here have tried to explain this field redefinition ambiguity more from a microscopic point of view, we have been completely agnostic about the microscopic descriptions or statistical interpretations of these field redefinitions. Consequently, we have found multiple ways of `integrating in' the non-fluid variables in the BDNK recast in the Landau frame. In principle, one could come with an infinite number of ways of performing this process.}.
\subsection{Convention and notations}
Throughout the manuscript, we have used natural unit ($\hbar = c = k_{B} = 1 $) and flat space-time with mostly positive metric signature $g^{\mu\nu} = \text{diag}\left(-1,1,1,1\right)$. $\varepsilon, T, P, u^{\mu}$ are respectively energy density, temperature pressure and hydrodynamic four-velocity. The local rest frame is defined as $u^{\mu}=(1,0,0,0)$, $\Delta^{\mu\nu}=g^{\mu\nu}+u^{\mu}u^{\nu}$ is the space projection operator orthogonal to $u^{\mu}$. $\Delta^{\mu\nu\alpha\beta}=\frac{1}{2}\Delta^{\mu\alpha}\Delta^{\nu\beta}+\frac{1}{2}\Delta^{\mu\beta}\Delta^{\nu\alpha}
-\frac{1}{3}\Delta^{\mu\nu}\Delta^{\alpha\beta}$ is the traceless projection operator orthogonal to $u_{\mu}$ and $\Delta_{\mu\nu}$. Any rank-2, symmetric, traceless tensor is defined as, $A^{\langle\mu}B^{\nu\rangle}=\Delta^{\mu\nu}_{\alpha\beta}A^{\alpha}B^{\beta}$.
The used derivative operators read as: covariant time derivative $D=u^{\mu}\partial_{\mu}$, spatial gradient $\nabla^{\mu}=\Delta^{\mu\nu}\partial_{\nu}$ and traceless, symmetric velocity gradient
$\sigma^{\mu\nu}=\partial^{\langle\mu}u^{\nu\rangle}$.
$\eta$ is the shear viscous coefficient, $\tau_{\pi}$ is the relaxation time of shear-viscous flow $\pi^{\mu\nu}$ of MIS theory, ${\chi},\theta$ are the first order field correction coefficients of BDNK theory. From the constraints of the second law of thermodynamics, $\eta$ should always be a positive number
\cite{Weinberg:1971mx}. The scaling notation $\tilde{x}$ denotes $x/(\varepsilon+P)$.
We linearize the conservation equations for small perturbations of fluid variables around their hydrostatic equilibrium,
 $\psi(t,x)=\psi_0+\delta\psi(t,x)$, with the perturbations expressed in the plane wave solutions via a Fourier transformation $\delta\psi(t,x)\rightarrow e^{i(kx-\omega t)} \delta\psi(\omega,k)$,
 (subscript $0$ indicates global equilibrium).
\subsection{Organization of the paper}
This note is organized as follows. In section \ref{section MIS}, we describe the MIS theory in its simplest form, and then we show how integrating out the extra `non-fluid' variable results in a stress tensor with an infinite number of derivatives. This section will act as a warm-up for the techniques of infinite sum to be used in the next section. Also, it indicates how a causal theory in the Landau frame, if expressed only in terms of fluid variables, turns out to have an infinite number of derivatives. In the next section \ref{section BDNK}, we describe the BDNK theory and redefine the velocity and temperature variables (only at the linearized level) to bring them to the Landau frame. Redefinition involves generating an infinite number of derivatives. We can sum these infinite series in two different ways as described in two different subsections of section \ref{section BDNK}. These two different ways of summation lead to two different methods of `integrating in' new `non-fluid' variables, showing the non-uniqueness of the process of `integrating in' new variables. In section \ref{section-dis}, the dispersion relations and the corresponding spectra of these different systems of equations have been analyzed to check that our systems of equations are indeed
equivalent to BDNK formalism, at least in the hydrodynamic regime. Finally, in section \ref{section Conclude}, we conclude.

\section{MIS theory - an infinite order fluid formalism}
\label{section MIS}
The pathologies regarding superluminal signal propagation and thermodynamic stability of the long-established relativistic first-order theories \cite{Landau,Eckart:1940te}, have been first taken care by the higher order MIS theory \cite{Hiscock:1983zz,Hiscock:1985zz}, where the dissipative field corrections are promoted to new degrees of freedom \cite{Muller,Israel:1976tn,Israel:1976efz,Israel:1979wp}. Keeping up to the linear terms, the MIS equations of motion are given by
\cite{Romatschke:2009im},
\begin{align}
&\partial_{\mu}T^{\mu\nu}=0~,~~T^{\mu\nu}=\varepsilon\left[u^{\mu}u^{\nu}+\frac{1}{3}\Delta^{\mu\nu}\right]+\pi^{\mu\nu}~,
\label{MIS1}\\
&\pi^{\mu\nu}+\tau_{\pi}D\pi^{\mu\nu}=-2\eta\sigma^{\mu\nu} ~.
\label{MIS2}
\end{align}
Here, we attempt to derive the combined results of Eq.\eqref{MIS1} and \eqref{MIS2} without treating $\pi^{\mu\nu}$ as an independent degree of freedom. Instead of attributing an individual differential equation to $\pi^{\mu\nu}$ like Eq.\eqref{MIS2}, we express it as a sum of gradient corrections that includes all derivative orders in Eq.\eqref{MIS1} itself such as,
\begin{align}
&\pi^{\mu\nu}=\sum_{n=1}^{\infty}\pi_n^{\mu\nu}~,\nn
&\pi^{\mu\nu}_1=-2\eta\sigma^{\mu\nu}~,~~~\pi_n^{\mu\nu}=-\tau_{\pi}D\pi^{\mu\nu}_{n-1}~,n \geq 2~.
\label{pisum}
\end{align}
This leads to the shear-stress tensor as the following,
\begin{align}
 \pi^{\mu\nu}=&-2\eta\left\{\sum_{n=0}^{\infty}\left(-\tau_{\pi}D\right)^n\right\}\sigma^{\mu\nu}
 \label{MIS3}\\
 =&-2\eta \left(1+\tau_{\pi}D\right)^{-1}\sigma^{\mu\nu}~.
 \label{MIS3a}
\end{align}
So, we conclude that if we want to write the MIS theory without introducing any additional degrees of freedom, this will lead to a stress tensor that is defined up to all orders of gradient correction. Any finite truncation of Eq.\eqref{MIS3} fails to produce the relaxation operator like structure in the denominator of Eq.\eqref{MIS3a}. However, it is to be noted that Eq.\eqref{MIS2} is local in both time and space, whereas Eq.\eqref{MIS3a} becomes non-local in time since the frequency of the corresponding Fourier mode appears in the denominator. The details of the acausality of a truncated series in Eq.\eqref{MIS3} can be found in \cite{Mitra:2023ipl}.
\section{BDNK theory and the transformation of the `fluid frame'}\label{section BDNK}
In the last few years, a new study of the relativistic first-order stable-causal theory (BDNK theory) has been proposed by defining the out-of-equilibrium hydrodynamic variables in a general frame other than that is defined by Landau or Eckart, through their postulated constitutive relations that include spatial as well as temporal gradients~\cite{Bemfica:2017wps,Bemfica:2019knx,Bemfica:2020zjp,Kovtun:2019hdm,Hoult:2020eho,Hoult:2021gnb} . In BDNK theory, if we further impose conformal symmetry and no conserved charges, the energy-momentum tensor ($T^{\mu\nu}$) takes the form,
\begin{align}
T^{\mu\nu}&=(\varepsilon +{\cal{A}})\left[u^{\mu}u^{\nu}+\frac{\Delta^{\mu\nu}}{3}\right]\nn
&+\left[u^{\mu}Q^{\nu}+u^{\nu}Q^{\mu}\right]-2\eta\sigma^{\mu\nu},
\label{BDNK1}
\end{align}
with the first-order dissipative field corrections as,
\begin{align}
{\cal{A}}={\chi}\left[3\frac{DT}{T}+\nabla_{\mu}u^{\mu}\right],~~
Q^{\mu}=\theta\left[\frac{\nabla^{\mu}T}{T}+Du^{\mu}\right].
 \label{BDNK2}
 \end{align}
We have used the identity $D\varepsilon/(\varepsilon+P)=3DT/T$ for a conformal system where the energy density goes as $\varepsilon\sim T^4$ and it is connected with the pressure $P$ as $(\varepsilon+P)=4\varepsilon/3$ \footnote{For simplicity, throughout this note, we shall restrict our analysis to conformal fluids, where temperature provides the only scale and the space-time dependence of all other dimensional variables like energy density is completely determined by that of the temperature. For example, $\varepsilon (x^\mu) = 3c~T^4(x^\mu),~~~P(x^\mu) = c~T^4(x^\mu),~~\text{where $c$ is some constant.}$ Because of this, while discussing the space-time dependence of the fluid variables, we shall often use $\varepsilon(x^\mu)$, $P(x^\mu)$ or $T(x^\mu)$ interchangeably.}. The dispersion relations resulting from Eq.\eqref{BDNK1} produce stable-causal modes only with non-zero values of $\theta$ and $\chi$. The neatness of this method lies in not requiring any additional degrees of freedom other than the temperature and velocity to preserve causality and stability.
Eq.\eqref{BDNK1} and \eqref{BDNK2} also show that the theory is local in fluid variables both spatially and temporally. However, as already discussed, the velocity or the temperature of the fluid is not defined here in terms of any microscopic operator like the stress tensor. In this section, we would like to redefine the velocity and the temperature in a way so that the stress tensor, expressed in terms of these redefined fluid variables, satisfies the Landau frame condition. Our philosophy is as follows.

We shall assume that the one-point function of the microscopic stress tensor operator in a `near thermal' state is given by the BDNK stress tensor \eqref{BDNK1}. But it is expressed in terms of some `velocity' and `temperature' variables $\{u^\mu, T\}$, which agree with the traditional definitions of velocity and temperature in global equilibrium but deviate in a generic `near equilibrium' state. On the other hand, we know that in the Landau frame, the velocity and the temperature fields
are locally defined in terms of the one-point function of the Stress tensor $T^{\mu \nu}$ as the following,
\begin{equation}\label{globalLandau}
T^{\mu}_{\nu}(\hat{T},\hat{u}^{\mu}) ~\hat u^\nu =-\hat\varepsilon~\hat u^\mu ~.
\end{equation}
We denote the notation $\hat{}$ to indicate fields at the Landau frame such as
$\hat u^\nu$ is the velocity, $\hat\varepsilon$ is the local energy density and $\hat{T}$ is the temperature in the Landau frame.
Transforming the BDNK stress tensor in the Landau frame involves two steps. First, we have to solve for $\hat u^\mu$ and $\hat\varepsilon$ by solving equation \eqref{globalLandau}, where in 
place of $T^\mu_\nu$ we shall use the BDNK stress tensor \eqref{BDNK1}. The second step involves rewriting the BDNK stress tensor in terms of these new fluid variables $\hat u^\mu$ and $\hat \varepsilon$.

Generically, performing such a frame transformation in a non-perturbative manner is extremely cumbersome. But to make our analysis computationally tractable, we restrict it to linearized treatment. Physically, we are restricting our analysis only to those fluid states whose deviation from global equilibrium is of very small amplitude. Such perturbations are enough to decide the linear stability and the causality of the theory - the key motivation behind the BDNK formalism. Since all definitions of the fluid variables agree in global equilibrium (or at the level of `ideal' fluid),  field redefinition is needed only in `non-equilibrium' fluid states. It follows that, if the deviation from equilibrium is of small amplitude such that a linearized treatment is allowed, the same should also be true for field redefinition. In other words, while redefining the velocity and the temperature, we can safely ignore terms that are nonlinear in the shift of the variables. In terms of equations, what we mean is the following.

We define that the velocity $u^\mu$ and the temperature $T$ in the BDNK stress tensor are related to the Landau frame velocity $\hat u^\mu$ and temperature $\hat T$ in the following fashion,
\begin{equation}\label{shift}
\begin{split}
u^\mu = \hat u^\mu +\delta u^\mu,~~~T = \hat T+ \delta T~,
\end{split}
\end{equation}
where the shift variables $\delta u^\mu$ and $\delta T$ are small enough to be treated only linearly. Note that both $\delta u^\mu$ and $\delta T$ are non-trivial functions of $\hat u^\mu$ and $\hat T$. Once we impose the Landau gauge condition \eqref{globalLandau} after substituting \eqref{shift} in the BDNK stress tensor \eqref{BDNK1}, it reduces to the following set of coupled and linear partial differential equations (PDEs) for the shift variables,
\begin{align}
 &\delta u^{\mu}+\tilde{\theta}\left[\hat{D}\hat{u}^{\mu}+\frac{\hat{\nabla}^{\mu}\hat{T}}{\hat{T}}\right]+\tilde{\theta}\left[\hat{D}\delta{u}^{\mu}+\frac{\hat{\nabla}^{\mu}\delta{T}}{\hat{T}}\right]=0,
 \label{ogv1}\\
&\frac{\delta T}{\hat{T}}+\tilde{\chi}\left[\frac{\hat{D}\hat{T}}{\hat{T}}
+ \frac{\hat{\nabla}_{\mu}\hat{u}^{\mu}}{3}\right]+\tilde{\chi}\left[\frac{\hat{D}\delta{T}}{\hat{T}}+\frac{\hat{\nabla}_{\mu}\delta{u}^{\mu}}{3}\right]=0~.
\label{ogT1}
\end{align}
This linearization simplifies the analysis so that we can have an `all-order' (in derivatives) formula for both the field redefinitions and the stress tensor in the new frame. 

It turns out that the `MIS type nonlocality' emerges here again, even in the BDNK theory, due to the infinite order field redefinition is needed to cast it in the Landau frame. At the linearized level, the field redefinition can be done in two different representations. In the first case, we have performed the infinite summation on only the temporal derivatives, which leads us to equations which are local in space, but are nonlocal in time due to the time derivative appearing in the denominator. In the second case, both the spatial and the temporal derivatives were summed, leading to a full nonlocal redefinition of the fluid variables. In either case, these nonlocalities (derivatives appearing in the denominator) could be absorbed by introducing new `non-fluid' variables. These two different methods are described in the following two different subsections.

\subsection{Method-1: Frame transformation order by order}\label{section BDNK!}

In this subsection, we shall solve these PDEs \eqref{ogv1} and \eqref{ogT1} order by order in derivative expansion. We shall assume that $\delta u^\mu, \delta \varepsilon$ and $\delta T$ admit the following infinite series expansion,
\begin{align}\label{expand}
 \delta u^\mu =\sum_{n=1}^{\infty}\delta u_n^{\mu},~~~\delta\varepsilon=\sum_{n=1}^{\infty}\delta\varepsilon_n,~~~\delta T =\sum_{n=1}^{\infty}\delta T_n~.
\end{align}
Here, the subscript $(n)$ denotes the order in terms of derivative expansion.
Substituting the expansion of \eqref{expand} in the PDEs \eqref{ogv1} and \eqref{ogT1}, one can easily find the solution in terms of the following recursive relations,
\begin{align}
&\delta T_1=-\tilde{\chi}\left[\frac{\hat{D} \hat{T}}{\hat{T}}+\frac{1}{3}\hat{\nabla}_{\mu}\hat{u}^{\mu}\right],~~~\delta u_1^{\mu}=-\tilde{\theta} \left[\frac{\hat{\nabla}^{\mu}\hat{T}}{\hat{T}}+\hat{D}\hat{u}^{\mu}\right],\nn
&\delta T_n=-\tilde{\chi}\left[\frac{1}{\hat{T}}\hat{D}\delta T_{n-1}+\frac{1}{3}\hat{\nabla}_{\mu}\delta u^{\mu}_{n-1}\right]~~~~\text{for}~n\geq 2~,
\nn
&\delta u_n^{\mu}=-\tilde{\theta} \left[\frac{1}{\hat{T}}\hat{\nabla}^{\mu}\delta T_{n-1}+\hat{D}\delta u^{\mu}_{n-1}\right]~~~~\text{for}~n\geq 2~.
\label{shift-soln}
\end{align}
Eq.\eqref{shift-soln} provides the successive field corrections up to any desired order.

The next step is to rewrite the energy-momentum tensor in terms of the new fluid variables.
The energy-momentum tensor in this frame turns out to be,

\begin{align}
T^{\mu\nu}=&\left[\hat{\varepsilon}+\sum_{n=1}^{\infty}\delta\varepsilon_n+\chi\left\{3\frac{\hat{D}\hat{T}}{\hat{T}}+\partial_{\alpha}\hat{u}^{\alpha}+\frac{3}{\hat{T}}\hat{D}\sum_{n=1}^{\infty}\delta T_n+\partial_{\alpha}\sum_{n=1}^{\infty}\delta u^{\alpha}_n\right\}\right]\left(\hat{u}^{\mu}\hat{u}^{\mu}+\frac{1}{3}\hat{\Delta}^{\mu\nu}\right)\nn
+&\left[\frac{4}{3}\hat{\varepsilon}\sum_{n=1}^{\infty}\delta u_n^{\nu}+\theta\left\{\frac{\hat{\nabla}^{\nu}\hat{T}}{\hat{T}}+\hat{D}\hat{u}^{\nu}+\frac{1}{\hat{T}}\hat{\nabla}^{\nu}\sum_{n=1}^{\infty}\delta T_n+\hat{D}\sum_{n=1}^{\infty}\delta u^{\nu}_n\right\}\right]\hat{u}^{\mu}\nn
+&\left[\frac{4}{3}\hat{\varepsilon}\sum_{n=1}^{\infty}\delta u_n^{\mu}+\theta\left\{\frac{\hat{\nabla}^{\mu}\hat{T}}{\hat{T}}+\hat{D}\hat{u}^{\mu}+\frac{1}{\hat{T}}\hat{\nabla}^{\mu}\sum_{n=1}^{\infty}\delta T_n+\hat{D}\sum_{n=1}^{\infty}\delta u^{\mu}_n\right\}\right]\hat{u}^{\nu}
-2\eta\left[\hat{\sigma}^{\mu\nu}+\sum_{n=1}^{\infty}\partial^{\langle\mu}\delta u_n^{\nu\rangle}\right] ~.
\label{BDNK3}
\end{align}

As mentioned before, only linearized terms are considered. The used notations (now defined in terms of Landau frame variable) read:
\begin{equation}
    \hat{\Delta}^{\mu\nu}=g^{\mu\nu}+\hat{u}^{\mu}\hat{u}^{\nu},~\hat{D}=\hat{u}^{\mu}\partial_{\mu},~\hat{\nabla}^{\mu}=\hat{\Delta}^{\mu\nu}\partial_{\nu},~\hat{\sigma}^{\mu\nu}=\partial^{\langle\mu}\hat{u}^{\nu\rangle}=\hat{\Delta}^{\mu\nu}_{\alpha\beta}\partial^{\alpha}\hat{u}^{\beta}
\end{equation}
After substituting the recursive solution for $\delta u^\mu_n$ and $\delta T_n$ as given in \eqref{shift-soln}, the energy density correction and energy-flux or momentum flow vanish as expected in the Landau frame, and one finally has the following energy-momentum tensor upto all order,
\begin{align}
T^{\mu\nu}=&\hat{\varepsilon}\left[\hat{u}^{\mu}\hat{u}^{\mu}+\frac{1}{3}\hat{\Delta}^{\mu\nu}\right]
-2\eta\left[\hat{\sigma}^{\mu\nu}+\sum_{n=1}^{\infty}\partial^{\langle\mu}\delta u_n^{\nu\rangle}\right]~.
\label{BDNK4}
\end{align}

\subsubsection{All order sum of the temporal derivatives}\label{section BDNK2}
\label{allorder}
Once we explicitly evaluate $\delta u^\mu_n$ and $\delta T_n$ for the first few orders, we observe that a very nice pattern emerges, which we could use to sum this infinite series to get an all-order expression.

In order to do so, first we list the velocity and temperature corrections up to first four orders obtained from the Landau matching conditions:

\begin{align}
\delta u^{\mu}_1&=-\tilde{\theta}\left[\hat{D}\hat{u}^{\mu}+\frac{\hat{\nabla}^{\mu}\hat{T}}{\hat{T}}\right]~,
\label{vel1}\\
\frac{\delta T_1}{\hat{T}}&=-\tilde{\chi}\left[\frac{\hat{D}\hat{T}}{\hat{T}}+\frac{1}{3}\left(\hat{\nabla}\cdot\hat{u}\right)\right]
\label{temp1}~,\\
\delta u^{\mu}_2&=\tilde{\theta}^2\hat{D}^2
\hat{u}^{\mu}+\tilde{\theta}\left[\tilde{\theta}+\tilde{\chi}\right]\frac{1}{\hat{T}}\hat{D}\hat{\nabla}^{\mu}\hat{T}+\tilde{\theta}\frac{\tilde{\chi}}{3}\hat{\nabla}^{\mu}\left(\hat{\nabla}\cdot\hat{u}\right)~,
\label{vel2}\\
\frac{\delta T_2}{\hat{T}}&=\tilde{\chi}^2\frac{\hat{D}^2\hat{T}}{\hat{T}}+\frac{\tilde{\chi}}{3}\left[\tilde{\chi}+\tilde{\theta}\right]\hat{D}\left(\hat{\nabla}\cdot\hat{u}\right)+\frac{\tilde{\chi}}{3}\tilde{\theta}\frac{\hat{\nabla}^2\hat{T}}{\hat{T}}~,
\label{temp2}\\
\delta u^{\mu}_3&=-\tilde{\theta}^3\hat{D}^3\hat{u}^{\mu}-\tilde{\theta}\left[\tilde{\theta}^2+\tilde{\theta}\tilde{\chi}+\tilde{\chi}^2\right]\frac{\hat{D}^2\hat{\nabla}^{\mu}\hat{T}}{\hat{T}}
-\tilde{\theta}\frac{\tilde{\chi}}{3}\left[2\tilde{\theta}+\tilde{\chi}\right]\hat{D}\hat{\nabla}^{\mu}\left(\hat{\nabla}\cdot\hat{u}\right)-\tilde{\theta}^2\frac{\tilde{\chi}}{3}\frac{\hat{\nabla}^2\hat{\nabla}^{\mu}\hat{T}}{\hat{T}},
\label{vel3}\\
\frac{\delta T_3}{\hat{T}}&=-\tilde{\chi}^3\frac{\hat{D}^3\hat{T}}{\hat{T}}-\frac{\tilde{\chi}}{3}\left[\tilde{\chi}^2+\tilde{\chi}\tilde{\theta}+\tilde{\theta}^2\right]\hat{D}^2\left(\hat{\nabla}\cdot\hat{u}\right)
-\frac{\tilde{\chi}}{3}\tilde{\theta}\left[2\tilde{\chi}+\tilde{\theta}\right]\frac{\hat{D}\hat{\nabla}^2\hat{T}}{\hat{T}}-\frac{\tilde{\chi}^2}{9}\tilde{\theta}\hat{\nabla}^2\left(\hat{\nabla}\cdot\hat{u}\right)~,
\label{temp3}\\
\delta u^{\mu}_4&=\tilde{\theta}^4\hat{D}^4\hat{u}^{\mu}+\tilde{\theta}\left[\tilde{\theta}^3+\tilde{\theta}^2\tilde{\chi}+\tilde{\theta}\tilde{\chi}^2+\tilde{\chi}^3\right]\frac{\hat{D}^3\hat{\nabla}^{\mu}\hat{T}}{\hat{T}}
+\tilde{\theta}\frac{\tilde{\chi}}{3}\left[3\tilde{\theta}^2+2\tilde{\theta}\tilde{\chi}+\tilde{\chi}^2\right]\hat{D}^2\hat{\nabla}^{\mu}\left(\hat{\nabla}\cdot\hat{u}\right)\nn
&+\tilde{\theta}^2\frac{\tilde{\chi}}{3}\left[2\tilde{\theta}+2\tilde{\chi}\right]\hat{D}\frac{\hat{\nabla}^2\hat{\nabla}^{\mu}\hat{T}}{\hat{T}}+
\tilde{\theta}^2\frac{\tilde{\chi}^2}{9}\hat{\nabla}^2\hat{\nabla}^{\mu}\left(\hat{\nabla}\cdot\hat{u}\right),
\label{vel4}\\
\frac{\delta T_4}{\hat{T}}&=\tilde{\chi}^4\frac{\hat{D}^4\hat{T}}{\hat{T}}+\frac{\tilde{\chi}}{3}\left[\tilde{\chi}^3+\tilde{\chi}^2\tilde{\theta}+\tilde{\chi}\tilde{\theta}^2+\tilde{\theta}^3\right]\hat{D}^3\left(\hat{\nabla}\cdot\hat{u}\right)
+\frac{\tilde{\chi}}{3}\tilde{\theta}\left[3\tilde{\chi}^2+2\tilde{\chi}\tilde{\theta}+\tilde{\theta}^2\right]\frac{\hat{D}^2\hat{\nabla}^2\hat{T}}{\hat{T}}
\nn
&+\frac{\tilde{\chi}^2}{9}\tilde{\theta}\left[2\tilde{\chi}+2\tilde{\theta}\right]\hat{D}\hat{\nabla}^2\left(\hat{\nabla}\cdot\hat{u}\right)+\frac{\tilde{\chi}^2}{9}\tilde{\theta}^2\frac{\hat{\nabla}^4\hat{T}}{\hat{T}}~,\\
\label{temp4}
&\vdots\nonumber~~~~~.
\end{align}

We see that, with increasing order $n$ of derivative correction, the velocity correction terms $\delta u_n^{\mu}$ (as well as the temperature correction terms $\delta T_n$),
include higher and higher orders of the spatial gradients on $T$ and $u^{\mu}$ systematically. Furthermore, the order of the time derivative on each such spatial derivative term successively increases. This increase of the temporal gradients follows a particular pattern so that they can be combined into products of infinite sums.
Below, we write the fully summed (up to all orders) velocity and the temperature corrections such that this repetitive pattern in the temporal derivatives becomes manifest.

\begin{align}
u^{\mu}&=\hat{u}^{\mu}+\delta u_1^{\mu}+\delta u_2^{\mu}+\cdots=\left[1+\left(-\tilde{\theta}\hat{D}\right)+\left(-\tilde{\theta}\hat{D}\right)^2+\left(-\tilde{\theta}\hat{D}\right)^3+\cdots
\right]\hat{u}^{\mu}\nonumber\\
&+(-\tilde{\theta})\left[1+\left(-\tilde{\theta}\hat{D}\right)+\left(-\tilde{\theta}\hat{D}\right)^2+\cdots
\right]\left[1+\left(-\tilde{\chi}\hat{D}\right)+\left(-\tilde{\chi}\hat{D}\right)^2+\cdots\right]\frac{\hat{\nabla}^{\mu}\hat{T}}{\hat{T}}\nonumber\\
&+(-\tilde{\theta})\left(-\frac{\tilde{\chi}}{3}\right))
\left[1+2\left(-\tilde{\theta}\hat{D}\right)+3\left(-\tilde{\theta}\hat{D}\right)^2+\cdots\right]\left[1+\left(-\tilde{\chi}\hat{D}\right)+\left(-\tilde{\chi}\hat{D}\right)^2+\cdots\right]
\hat{\nabla}^{\mu}\left(\hat{\nabla}\cdot\hat{u}\right)\nonumber\\
&+(-\tilde{\theta})^2\left(-\frac{\tilde{\chi}}{3}\right)
\left[1+2\left(-\tilde{\theta}\hat{D}\right)+3\left(-\tilde{\theta}\hat{D}\right)^2+\cdots\right]\left[1+2\left(-\tilde{\chi}\hat{D}\right)+3\left(-\tilde{\chi}\hat{D}\right)^2+\cdots\right]
\frac{\hat{\nabla}^2\hat{\nabla}^{\mu}\hat{T}}{\hat{T}}\nonumber\\
&+(-\tilde{\theta})^2\left(-\frac{\tilde{\chi}}{3}\right)^2
\left[1+3\left(-\tilde{\theta}\hat{D}\right)+6\left(-\tilde{\theta}\hat{D}\right)^2+\cdots\right]\left[1+2\left(-\tilde{\chi}\hat{D}\right)+3\left(-\tilde{\chi}\hat{D}\right)^2+\cdots\right]\hat{\nabla}^2
\hat{\nabla}^{\mu}\left(\hat{\nabla}\cdot\hat{u}\right)\nonumber\\
&+\cdots~~~~~.
\end{align}
  
The infinite sums over the time derivative can be encompassed in a closed form following the relaxation operator-like terms to appear in the denominator of the thermodynamic quantities, giving rise to pole-like structures in the following manner,
\begin{align}\label{resumvel}
u^{\mu}=&\hat{u}^{\mu}+\delta u_1^{\mu}+\delta u_2^{\mu}+\cdots\nn
=&\frac{1}{(1+\tilde{\theta}\hat{D})}\hat{u}^{\mu}\nonumber\\
+&(-\tilde{\theta})\frac{1}{(1+\tilde{\theta}\hat{D})}\frac{1}{(1+\tilde{\chi}\hat{D})}\frac{\hat{\nabla}^{\mu}\hat{T}}{\hat{T}}\nn
+&(-\tilde{\theta})\left(-\frac{\tilde{\chi}}{3}\right)\frac{1}{(1+\tilde{\theta}\hat{D})^2}\frac{1}{(1+\tilde{\chi}\hat{D})}\hat{\nabla}^{\mu}\hat{\nabla}\cdot\hat{u}\nonumber\\
+&(-\tilde{\theta})^2\left(-\frac{\tilde{\chi}}{3}\right)\frac{1}{(1+\tilde{\theta}\hat{D})^2}\frac{1}{(1+\tilde{\chi}\hat{D})^2}
\frac{\hat{\nabla}^2\hat{\nabla}^{\mu}\hat{T}}{\hat{T}}\nonumber\\
+&(-\tilde{\theta})^2\left(-\frac{\tilde{\chi}}{3}\right)^2\frac{1}{(1+\tilde{\theta}\hat{D})^3}\frac{1}{(1+\tilde{\chi}\hat{D})^2}
\hat{\nabla}^2\hat{\nabla}^{\mu}\hat{\nabla}\cdot\hat{u}\nonumber\\
+&\cdots~~~~~.
\end{align}

Similarly, for the temperature correction, we have the following derivative pattern :
 
\begin{align}\label{temppat}
T&=\hat{T}+\delta T_1+\delta T_2+\cdots=\left[1+\left(-\tilde{\chi}\hat{D}\right)+\left(-\tilde{\chi}\hat{D}\right)^2+\left(-\tilde{\chi}\hat{D}\right)^3+\cdots\right]\hat{T}\nonumber\\
&+\hat{T}(-\frac{\tilde{\chi}}{3})\left[1+\left(-\tilde{\chi}\hat{D}\right)+\left(-\tilde{\chi}\hat{D}\right)^2+\cdots\right]\left[1+\left(-\tilde{\theta}\hat{D}\right)+\left(-\tilde{\theta}\hat{D}\right)^2+\cdots\right]\left(\hat{\nabla}\cdot\hat{u}\right)\nonumber\\
&+\left(-\frac{\tilde{\chi}}{3}\right)(-\tilde{\theta})
\left[1+2\left(-\tilde{\chi}\hat{D}\right)+3\left(-\tilde{\chi}\hat{D}\right)^2+\cdots\right]\left[1+\left(-\tilde{\theta}\hat{D}\right)+\left(-\tilde{\theta}\hat{D}\right)^2+\cdots\right]
\hat{\nabla}^2\hat{T}\nn
&+\hat{T}\left(-\frac{\tilde{\chi}}{3}\right)^2(-\tilde{\theta})
\left[1+2\left(-\tilde{\chi}\hat{D}\right)+3\left(-\tilde{\chi}\hat{D}\right)^2+\cdots\right]
\left[1+2\left(-\tilde{\theta}\hat{D}\right)+3\left(-\tilde{\theta}\hat{D}\right)^2+\cdots\right]
\hat{\nabla}^2\left(\hat{\nabla}\cdot\hat{u}\right)\nn
&+\cdots ~~~~.
\end{align}
  
Just like the velocity variable, the above series can also be resummed as,
\begin{align}\label{tempresum}
T=&\hat{T}+\delta T_1+\delta T_2+\cdots\nn
=&\frac{1}{(1+\tilde{\chi}\hat{D})}\hat{T}\nonumber\\
+&\hat{T}\left(-\frac{\tilde{\chi}}{3}\right)\frac{1}{(1+\tilde{\chi}\hat{D})}\frac{1}{(1+\tilde{\theta}\hat{D})}\left(\hat{\nabla}\cdot\hat{u}\right)\nn
+&\left(-\frac{\tilde{\chi}}{3}\right)(-\tilde{\theta})\frac{1}{(1+\tilde{\chi}\hat{D})^2}\frac{1}{(1+\tilde{\theta}\hat{D})}\hat{\nabla}^2\hat{T}\nonumber\\
+&\hat{T}\left(-\frac{\tilde{\chi}}{3}\right)^2(-\tilde{\theta})\frac{1}{(1+\tilde{\chi}\hat{D})^2}\frac{1}{(1+\tilde{\theta}\hat{D})^2}
\hat{\nabla}^2\left(\hat{\nabla}\cdot\hat{u}\right)\nonumber\\
+&\cdots~~~.
\end{align}

Putting the velocity correction given by Eq.\eqref{resumvel} in Eq.\eqref{BDNK4} we have the all order frame transformed BDNK stress tensor in Landau frame as,
\begin{align}
T^{\mu\nu}=\hat{\varepsilon}\left[\hat{u}^{\mu}\hat{u}^{\nu}+\frac{1}{3}\hat{\Delta}^{\mu\nu}\right]+\hat{\pi}^{\mu\nu}~,
\label{BDNK00}
\end{align}
with the shear stress $\hat{\pi}^{\mu\nu}=-2\eta\left[\hat{\sigma}^{\mu\nu}+\sum_{n=1}^{\infty}\partial^{\langle\mu}\delta u_n^{\nu\rangle}\right]$ as the only dissipative contribution, now resummed under the all order frame transformation as the following,
\begin{align}
\hat{\pi}^{\mu\nu}=&-2\eta\bigg[
\frac{\hat{\nabla}^{\langle\mu}\hat{u}^{\nu\rangle}}{(1+\tilde{\theta}\hat{D})}
+\frac{(-\tilde{\theta})}{(1+\tilde{\theta}\hat{D})}\frac{\frac{1}{\hat{T}}\hat{\nabla}^{\langle\mu}\hat{\nabla}^{\nu\rangle}\hat{T}}{(1+\tilde{\chi}\hat{D})}\nn
+&\frac{(-\tilde{\theta})}{(1+\tilde{\theta}\hat{D})^2}\frac{(-\frac{1}{3}\tilde{\chi})}{\left(1+\tilde{\chi}\hat{D}\right)}\hat{\nabla}^{\langle\mu}\hat{\nabla}^{\nu\rangle}\hat{\nabla}\cdot\hat{u}\nn
+&\frac{(-\tilde{\theta})^2}{(1+\tilde{\theta}\hat{D})^2}\frac{\left(-\frac{1}{3}\tilde{\chi}\right)}{(1+\tilde{\chi}\hat{D})^2}
\frac{1}{\hat{T}}\hat{\nabla}^{\langle\mu}\hat{\nabla}^{\nu\rangle}\hat{\nabla}^2\hat{T}\nn
+&\frac{(-\tilde{\theta})^2}{(1+\tilde{\theta}\hat{D})^3}\frac{\left(-\frac{1}{3}\tilde{\chi}\right)^2}{(1+\tilde{\chi}\hat{D})^2}
\hat{\nabla}^{\langle\mu}\hat{\nabla}^{\nu\rangle}\hat{\nabla}^2\hat{\nabla}\cdot\hat{u}+\cdots
\bigg].
\label{BDNK11}
\end{align}
Another point to notice here is that for each spatial derivative, the time derivative due to the infinite sum also increases in the denominator in such a way that they exactly balance each other. Following the analysis of \cite{Hoult:2023clg}, this is a necessary condition for causality.

Equations like \eqref{resumvel} and \eqref{tempresum} are just formal solutions and make more sense in the frequency space than in the space of real-time due to the presence of derivatives in the denominator, which is an indication of nonlocality (or integration in real-time).
Just like in the MIS theory, such nonlocalities could be recast into a local set of equations by introducing new `non-fluid' variables, which is the topic of the next subsection.

\subsubsection{Introducing `non-fluid' degrees of freedom to make BDNK a local theory in Landau frame}

In section \ref{allorder}, Eq.\eqref{BDNK00} and \eqref{BDNK11}
combinedly provide the energy-momentum tensor of a frame-transformed BDNK theory that is nonlocal in fluid variables. In this subsection, our goal is to introduce new `non-fluid' degrees of freedom, ones that vanish at any state of global thermal equilibrium and, therefore, are not extensions of any conserved charges. This viewpoint also provides us some guidance as to how we should formulate the equations of motion for `non-fluid' variables. Like the $\pi^{\mu\nu}$ degrees of freedom in the MIS theory, any `non-fluid' degrees of freedom in the theory should tend to 0 in `relaxation type' equations as time progresses. The infinite sum of the temporal derivatives performed in the previous subsection provides the relaxation timescales via the poles involved in the summations. But here, unlike the MIS theory, after the infinite summation of the temporal derivatives, the degree of the pole keeps on increasing with an increasing number of spatial derivatives in the numerator ad infinitum. This leads to an infinite tower of non-fluid degrees of freedom in a nested series of `relaxation equations.'

We can make this intuition precise in the following set of infinitely many equations.
This is a local theory both in space and time, equivalent to BDNK, at least with respect to linearized perturbations around equilibrium in the hydrodynamic regime (barring a few singular points in the frequency domain), but has an infinite number of degrees of freedom, (as we expected) in the following manner,
\begin{align}
 &\partial_\mu T^{\mu\nu}=0~,~~~~~~~~
 T^{\mu\nu}=\hat{\varepsilon}\left[\hat{u}^{\mu}\hat{u}^{\nu}+\frac{1}{3}\hat{\Delta}^{\mu\nu}\right]+\hat{\pi}^{\mu\nu},\nn
 &(1+\tilde\theta\hat D)\hat{\pi}^{\mu\nu}=-2\eta\hat{\sigma}^{\mu\nu}+\rho_1^{\mu\nu},\nn
 &(1+\tilde\chi\hat D)\rho^{\mu\nu}_1=(-2\eta)(-\tilde{\theta})\frac{1}{\hat{T}}\hat{\nabla}^{\langle\mu}\hat{\nabla}^{\rangle\nu} \hat{T}+\rho_2^{\mu\nu},\nn
 &(1+\tilde\theta\hat D)\rho^{\mu\nu}_2=(-2\eta)(-\tilde{\theta})\left(-\frac{\tilde{\chi}}{3}\right)\hat{\nabla}^{\langle\mu}\hat{\nabla}^{\rangle\nu} \hat{\nabla}\cdot\hat{u}+\rho_3^{\mu\nu},\nn
 &(1+\tilde\chi\hat D)\rho^{\mu\nu}_3=(-2\eta)(-\tilde{\theta})^2\left(-\frac{\tilde{\chi}}{3}\right)\frac{1}{\hat{T}}\hat{\nabla}^{\langle\mu}\hat{\nabla}^{\rangle\nu} \hat{\nabla}^2 \hat{T}+\rho_4^{\mu\nu},\nn
 &(1+\tilde\theta\hat D)\rho^{\mu\nu}_4=(-2\eta)(-\tilde{\theta})^2\left(-\frac{\tilde{\chi}}{3}\right)^2\hat{\nabla}^{\langle\mu}\hat{\nabla}^{\rangle\nu}\hat{\nabla}^2\hat{\nabla}\cdot\hat{u}+\cdots\nn
 &\vdots~~~~~~
 \label{MIS-type1}
\end{align}
Eq. \eqref{MIS-type1} and so on set an infinite nested series of new degrees of freedom much in the same line as the conventional MIS theory given by Eq.\eqref{MIS1} and \eqref{MIS2}. Eq.\eqref{MIS-type1} combinedly boils down to Eq.\eqref{BDNK00} and \eqref{BDNK11} where each increasing spatial gradient term is now attributed to a new degree of freedom.

\subsection{Method-2: Frame transformation in one go}
In the previous section, we have solved the linearized frame transformation equations
\eqref{ogv1} and \eqref{ogT1} using derivative expansion. Though the method of derivative expansion could be applied to solve even a nonlinear set of equations, we have heavily used linearization to simplify the solution further.  In fact, the way we have summed the infinite series to generate temporal derivatives in the denominator is clearly a formal manipulation, and it makes sense only in the case of linearized treatment in Fourier space. It also indicates an integration over time, which is then made local by introducing new `non-fluid' variables.

Now, while solving \eqref{ogv1} and \eqref{ogT1}, if we eventually allow ourselves to have temporal derivatives ($\hat D$) in the denominator, there is no harm in having spatial derivatives as well (again makes sense only when viewed in Fourier space and indicates an infinite order of spatial derivatives or integration/nonlocality in space). In this subsection, we shall use this formal manipulation of having both spatial and temporal derivatives in the denominator. This will lead to solutions of the frame transformation equations \eqref{ogv1} and  \eqref{ogT1} in one go.

The steps are as follows. First, we take the divergence of equation \eqref{ogv1} and the following two coupled scalar equations will give the two scalar variables $(\hat\nabla\cdot \delta u)$ and $\delta T/\hat{T}$ as,
\begin{align}
&\left[1+\tilde{\theta}\hat{D}\right](\hat{\nabla}\cdot\delta u)+\tilde{\theta}\hat{\nabla}^2\frac{\delta{T}}{\hat{T}}+\tilde{\theta}\left[\frac{\hat{\nabla}^2\hat{T}}{\hat{T}}+\hat{D}\hat{\nabla}\cdot\hat{u}\right]=0~,
\label{ogc1}\\
&\left[1+\tilde{\chi}\hat{D}\right]\frac{\delta T}{\hat{T}}+\frac{\tilde{\chi}}{3}(\hat{\nabla}\cdot\delta u)=0~.
\label{ogc2}
\end{align}
In Eq.\eqref{ogc2} we have used the on shell identity $\frac{\hat{D}\hat{T}}{\hat{T}}+\frac{1}{3}\hat{\nabla}\cdot\hat{u}=0$ that always holds at linearized level under Landau frame condition.
Now eliminating $(\hat\nabla\cdot\delta u)$ from the above two equations, first we find $\delta T\over \hat T$. Then, substituting this solution in \eqref{ogv1}, we find the expression for $\delta u^\mu$. The final solution (BDNK variables in terms of Landau frame variables) takes the following form:
 
\begin{align}
u^\mu=\left(\hat{u}^{\mu}+\delta u^{\mu}\right)
&=\left(\hat{u}^{\mu}\over 1+\tilde{\theta}\hat{D}\right)\notag\\
&+\left(1+\tilde{\theta}\hat{D}\right)^{-1}\left[(1+\tilde{\theta}\hat{D})(1+\tilde{\chi}\hat{D})-\tilde{\theta}\frac{\tilde{\chi}}{3}~\hat{\nabla}^2\right]^{-1}\bigg[-\tilde{\theta}\frac{\hat{\nabla}^{\mu}\hat{T}}{\hat{T}}+\frac{\tilde{\theta}}{3}\left(\tilde{\theta}+\tilde{\chi}\right)\hat{\nabla}^{\mu}(\hat{\nabla}\cdot \hat{u})\bigg]~,
\label{ogv2}\\
T=\left(\hat{T}+\delta T\right)&=\left[(1+\tilde{\theta}\hat{D})(1+\tilde{\chi}\hat{D})-\tilde{\theta}\frac{\tilde{\chi}}{3}~\hat{\nabla}^2\right]^{-1}\bigg[(1+\tilde{\theta}\hat{D})\hat{T}-\frac{\tilde{\chi}}{3}\hat{T}\left(\hat{\nabla}\cdot\hat{u}\right)\bigg]~.
\label{ogt2}
\end{align}
  
In the Landau frame, the stress tensor will again have the structure of the form given in equation \eqref{BDNK4}. After substituting the solutions \eqref{ogv2} there, we finally get the following shear tensor,
  
 \begin{align}
\hat{\pi}^{\mu\nu}=&
-\left[2\eta\over1+\tilde{\theta}\hat{D}\right]\hat{\sigma}^{\mu\nu}+\left[2\eta\tilde{\theta}\over 1+\tilde{\theta}\hat{D}\right]\left[{\frac{\hat{\nabla}^{\langle\mu}\hat{\nabla}^{\nu\rangle}\hat{T}}{\hat{T}}-\frac{1}{3}(\tilde{\theta}+\tilde{\chi})\hat{\nabla}^{\langle\mu}\hat{\nabla}^{\nu\rangle}\left(\hat{\nabla}\cdot\hat{u}\right)\over (1+\tilde{\theta}\hat{D})(1+\tilde{\chi}\hat{D})-\tilde{\theta}\frac{\tilde{\chi}}{3}~\hat{\nabla}^2}\right]~.
\label{BDNK33}
\end{align}
  
Equation \eqref{BDNK33} could be further simplified using the fact that in Landau frame at the linearized level $\hat\nabla\cdot\hat u$ and $\hat D\hat T\over\hat  T$ are related as follows,
\be
\hat\nabla\cdot\hat u+3 \left(\hat D\hat T\over\hat  T\right) = \text{terms nonlinear in perturbations.}
\label{identity}
\ee
Using identity \eqref{identity}, Eq.\eqref{BDNK33} becomes,
\begin{align}
\hat{\pi}^{\mu\nu}&=
-\left[2\eta\over1+\tilde{\theta}\hat{D}\right]\hat{\sigma}^{\mu\nu}\nn
&+\left[2\eta\tilde{\theta}\over 1+\tilde{\theta}\hat{D}\right]\left[{\left\{1+(\tilde{\theta}+\tilde{\chi})\hat{D}\right\}\frac{\hat{\nabla}^{\langle\mu}\hat{\nabla}^{\nu\rangle}\hat{T}}{\hat{T}}\over (1+\tilde{\theta}\hat{D})(1+\tilde{\chi}\hat{D})-\frac{\tilde{\theta}\tilde{\chi}}{3}~\hat{\nabla}^2}\right]~.
\label{BDNK34}
\end{align}
The equations \eqref{ogv2}, \eqref{ogt2}, \eqref{BDNK33} and \eqref{BDNK34} are all very formal with spatial as well as temporal derivatives in the denominator. But following the strategy presented in the case of MIS theory, we can recast equation \eqref{BDNK34} as an inhomogeneous differential equation for the new `nonfluid' degree of freedom $\pi^{\mu\nu}$ as follows,
\begin{align}
&\left[(1+\tilde{\theta}\hat{D})(1+\tilde{\chi}\hat{D})-\tilde{\theta}\frac{\tilde{\chi}}{3}~\hat{\nabla}^2\right]
\left\{(1+\tilde{\theta}\hat{D})\hat{\pi}^{\mu\nu}+2\eta\hat{\sigma}^{\mu\nu}\right\}
\nn
&~~~~~~ =~
2\eta\tilde{\theta}\left\{1+(\tilde{\theta}+\tilde{\chi})\hat{D}\right\}\frac{\hat{\nabla}^{\langle\mu}\hat{\nabla}^{\nu\rangle}\hat{T}}{\hat{T}}~.
\label{BDNKeqn}
\end{align}
Here, just like in MIS theory, we are introducing only one new `non-fluid' tensorial degree of freedom, but it follows a complicated inhomogeneous PDE, second order in spatial but third order in temporal derivatives\footnote{Note that in the limit $\tilde \chi\rightarrow0$, the equation \eqref{BDNKeqn} becomes very similar to the corresponding equation in MIS theory with a slight modification as follows.
\begin{align}
&(1+\tilde{\theta}\hat{D})\hat{\pi}^{\mu\nu}=
-2\eta\left[\hat{\sigma}^{\mu\nu}-\tilde{\theta}\left(\frac{\hat{\nabla}^{\langle\mu}\hat{\nabla}^{\nu\rangle}\hat{T}}{\hat{T}}\right)\right]~.
\end{align}
}.

\subsubsection{Comparison with the previous method with infinite `non-fluid' variables}
Generically, a nonlocal theory could be made local by introducing new degrees of freedom, but the process of `integrating in' new degrees could have ambiguities.
The two methods described in the previous two subsections could be one example of this ambiguity. Both methods attempt to write a system of coupled equations involving both fluid and `non-fluid' variables that are equivalent to the equations in BDNK theory. 
However, the widely different structure of the equations and the extra `non-fluid' variables in the two cases leads us to introduce an infinite number of variables in the first case, but only one such `non-fluid' degree of freedom suffices in the second case.
In this subsection, we would like to see how these two sets of equations are actually equivalent, at least in some regime of frequency and spatial momenta.

It turns out that the field redefinition we have used in the first method (see equations \eqref{resumvel} and \eqref{tempresum}) could be further rearranged in the following fashion. For the velocity redefinition, we have,
 
\begin{align}\label{velrearrange}
u^{\mu}=&\hat{u}^{\mu}+\delta u_1^{\mu}+\delta u_2^{\mu}+\cdots\notag\\
=&\frac{1}{(1+\tilde{\theta}\hat{D})}\hat{u}^{\mu}
+\frac{(-\tilde{\theta})}{(1+\tilde{\theta}\hat{D})}\frac{1}{(1+\tilde{\chi}\hat{D})}\left[1+\frac{(-\tilde{\theta})}{(1+\tilde{\theta}\hat{D})}\frac{\left(-\frac{\tilde{\chi}}{3}\right)}{(1+\tilde{\chi}\hat{D})}\hat{\nabla}^2+\cdots\right]\frac{\hat{\nabla}^{\mu}\hat{T}}{\hat{T}}\nonumber\\
+&\frac{(-\tilde{\theta})}{(1+\tilde{\theta}\hat{D})^2}\frac{\left(-\frac{\tilde{\chi}}{3}\right)}{(1+\tilde{\chi}\hat{D})}\left[1+\frac{(-\tilde{\theta})}{(1+\tilde{\theta}\hat{D})}\frac{\left(-\frac{\tilde{\chi}}{3}\right)}{(1+\tilde{\chi}\hat{D})}\hat{\nabla}^2+\cdots\right]\hat{\nabla}^{\mu}\left(\hat{\nabla}\cdot\hat{u}\right).
\end{align}
  
Similarly, for the temperature redefinition we have,
 
\begin{align}\label{temprearrange}
T=\hat{T}+\delta T_1+\delta T_2+\cdots=&\frac{1}{(1+\tilde{\chi}\hat{D})}\left[1+\frac{(-\tilde{\theta})}{(1+\tilde{\theta}\hat{D})}\frac{\left(-\frac{\tilde{\chi}}{3}\right)}{(1+\tilde{\chi}\hat{D})}\hat{\nabla}^2+\cdots\right]{\hat{T}}
\nn
+&\hat{T}\frac{\left(-\frac{\tilde{\chi}}{3}\right)}{(1+\tilde{\chi}\hat{D})}\frac{1}{(1+\tilde{\theta}\hat{D})}\left[1+\frac{(-\tilde{\theta})}{(1+\tilde{\theta}\hat{D})}\frac{\left(-\frac{\tilde{\chi}}{3}\right)}{(1+\tilde{\chi}\hat{D})}\hat{\nabla}^2+\cdots\right]\left(\hat{\nabla}\cdot\hat{u}\right)~.
\end{align}
  
Substituting this rearranged field redefinition, the dissipative part of the stress tensor could also be rearranged as,
 
\begin{align}
 \hat{\pi}^{\alpha\beta}=&-2\eta\hat{\Delta}^{\alpha\beta}_{\mu\nu} \Bigg[\frac{1}{(1+\tilde{\theta}\hat{D})}\hat{\nabla}^{\mu}\hat{u}^{\nu}\nonumber\\
 &+\frac{(-\tilde{\theta})}{(1+\tilde{\theta}\hat{D})}\frac{1}{(1+\tilde{\chi}\hat{D})}\frac{1}{\hat{T}}\hat{\nabla}^{\mu}\hat{\nabla}^{\nu}\left\{1+\frac{(-\tilde{\theta})}{(1+\tilde{\theta}\hat{D})}\frac{(-\frac{1}{3}\tilde{\chi})}{(1+\tilde{\chi}\hat{D})}\hat{\nabla}^2+\cdots\right\}\hat{T}\nonumber\\
 &+\frac{(-\tilde{\theta})}{(1+\tilde{\theta}\hat{D})^2}\frac{(-\frac{1}{3}\chi)}{(1+\tilde{\chi}\hat{D})}\hat{\nabla}^{\mu}\hat{\nabla}^{\nu}\left\{1+\frac{(-\tilde{\theta})}{(1+\tilde{\theta}\hat{D})}\frac{(-\frac{1}{3}\tilde{\chi})}{(1+\tilde{\chi}\hat{D})}\hat{\nabla}^2+\cdots\right\}\hat{\nabla}_{\rho}\hat{u}^{\rho}\Bigg]~.
 \label{BDNK22}
\end{align}
  
Now the infinite sum in powers of spatial derivative $\hat\nabla^2$ converges for those linearized perturbations where the operator satisfies the inequality,
\be
\bigg[\frac{\left(\frac{\tilde\theta\tilde{\chi}}{3}\right)\hat{\nabla}^2}{(1+\tilde{\theta}\hat{D})(1+\tilde{\chi}\hat{D})}\bigg]<1~.
\ee
Within this radius of convergence, we can again sum the spatial derivatives and get the following expression for the field redefinitions,
 
\begin{align}
u^{\mu}=&\hat{u}^{\mu}+\delta u_1^{\mu}+\delta u_2^{\mu}+\cdots \notag\\
=&\frac{1}{(1+\tilde{\theta}\hat{D})}\hat{u}^{\mu}\nn
+&(-\tilde{\theta})
\frac{\frac{\hat{\nabla}^{\mu}\hat{T}}{\hat{T}}}{\left[(1+\tilde{\theta}\hat{D})(1+\tilde{\chi}\hat{D})-\tilde{\theta}\frac{\tilde{\chi}}{3}\hat{\nabla}^2\right]}
+\frac{(-\tilde{\theta})\left(-\frac{\tilde{\chi}}{3}\right)}{(1+\tilde{\theta}\hat{D})}\frac{\hat{\nabla}^{\mu}\left(\hat{\nabla}\cdot\hat{u}\right)}{\left[(1+\tilde{\theta}\hat{D})(1+\tilde{\chi}\hat{D})-\tilde{\theta}\frac{\tilde{\chi}}{3}\hat{\nabla}^2\right]}~,\nn
=&\frac{1}{(1+\tilde{\theta}\hat{D})}\hat{u}^{\mu}+\frac{\left[-\tilde{\theta}\frac{\hat{\nabla}^{\mu}\hat{T}}{\hat{T}}+\frac{\tilde{\theta}}{3}\left(\tilde{\theta}+\tilde{\chi}\right)\hat{\nabla}^{\mu}(\hat{\nabla}\cdot \hat{u})\right]}{(1+\tilde{\theta}\hat{D})\left[(1+\tilde{\theta}\hat{D})(1+\tilde{\chi}\hat{D})-\tilde{\theta}\frac{\tilde{\chi}}{3}\hat{\nabla}^2\right]}~,
\label{vspaceresum}
\end{align}
  
and,
 
\begin{align}
T&=\hat{T}+\delta T_1+\delta T_2+\cdots
\nn
&=\frac{(1+\tilde{\theta}\hat{D})\hat{T}}{\left[(1+\tilde{\theta}\hat{D})(1+\tilde{\chi}\hat{D})-\tilde{\theta}\frac{\tilde{\chi}}{3}\hat{\nabla}^2\right]}
+\hat{T}\left(-\frac{\tilde{\chi}}{3}\right)\frac{\left(\hat{\nabla}\cdot\hat{u}\right)}{\left[(1+\tilde{\theta}\hat{D})(1+\tilde{\chi}\hat{D})-\tilde{\theta}\frac{\tilde{\chi}}{3}\hat{\nabla}^2\right]}~.
\label{tspaceresum}
\end{align}
  
From \eqref{vspaceresum} it is simple to estimate $\pi^{\mu\nu}$ as,
 
\begin{align}
 \hat{\pi}^{\mu\nu}=&-2\eta\frac{\tilde{\sigma}^{\mu\nu}}{(1+\tilde{\theta}\hat{D})}-2\eta\frac{\left[-\tilde{\theta}~\frac{\hat{\nabla}^{\langle\mu}\hat{\nabla}^{\nu\rangle}\hat{T}}{\hat{T}}+\tilde{\theta}\frac{\tilde{\chi}}{3}\frac{1}{(1+\tilde{\theta}\hat{D})}\hat{\nabla}^{\langle\mu}\hat{\nabla}^{\nu\rangle}\left(\hat{\nabla}\cdot\hat{u}\right)\right]}{\left[(1+\tilde{\theta}\hat{D})(1+\tilde{\chi}\hat{D})-\tilde{\theta}\frac{\tilde{\chi}}{3}\hat{\nabla}^2\right]}~,\nn
 =&
-\left[2\eta\over1+\tilde{\theta}\hat{D}\right]\sigma^{\mu\nu}+\left[2\eta\tilde{\theta}\over 1+\tilde{\theta}\hat{D}\right]\left[{\left(1+(\tilde{\theta}+\tilde{\chi})\hat{D}\right)\left(\frac{\hat{\nabla}^{\langle\mu}\hat{\nabla}^{\nu\rangle}\hat{T}}{\hat{T}}\right)\over (1+\tilde{\theta}\hat{D})(1+\tilde{\chi}\hat{D})-\tilde{\theta}\frac{\tilde{\chi}}{3}~\hat{\nabla}^2}\right]~.
\label{pispaceresum}
 \end{align}
   
It can be observed that Eq.\eqref{vspaceresum}, \eqref{tspaceresum} and \eqref{pispaceresum} are exactly identical as \eqref{ogv2}, \eqref{ogt2} and \eqref{BDNK34} of the field correction at one go results. (In the second step of the derivation of \eqref{vspaceresum} and \eqref{pispaceresum}, we have taken recourse to the identity \eqref{identity}. For detailed steps of the summation, the reader may refer to appendix \ref{app2}.)
So, within the radius of convergence, both methods actually generate the same set of equations as expected.

At this stage, let us emphasize one point. This method of `integrating in' new `non-fluid' degrees of freedom with new equations of motion is highly non-unique, even at the linearized level. For example, we could have chosen $\delta u^\mu$ and $\delta T$ themselves to be the new `non-fluid' variables, satisfying the new equations as given in \eqref{ogv1} and \eqref{ogT1} and we could take a viewpoint that the $u^\mu$ and the $T$ fields in the BDNK theory are actually the Landau frame fluid variables plus `non-fluid' variables $\{\delta u^\mu, \delta T\}$. Note that though $\delta u^\mu$ and $\delta T$ would look very much like velocity and temperature corrections, they are still `non-fluid' variables in the Landau frame since they vanish in global equilibrium. Another choice of introducing infinitely many `non-fluid' degrees of freedom would be to simply use $\delta u^\mu_n$ and $\delta T_n$ (as defined in \eqref{expand})  and then the recursive equations \eqref{shift-soln} would turn out to be the new equations of motion.

The two choices of new variables, discussed here in detail, are basically guided by our sense of mathematical aesthetics and an attempt to adhere to the philosophy of MIS theory where the new  `non-fluid' variable is a rank-2 symmetric tensor, structurally very similar to the energy-momentum tensor. At the moment, we do not have any further physical support behind our choice of variables.

\section{Dispersion relation}
\label{section-dis}
As we have seen in the previous sections, a system of fluid equations with terms up to all orders in derivative expansion could be converted to PDEs with a finite number of derivatives, provided we introduce new `non-fluid' degrees of freedom. The `non-fluid' variables we introduced basically capture the effect of a formal infinite sum over derivatives, leading to pole-like structures in the momentum-frequency space.

Now, these infinite series in derivatives (or, more precisely, in the 4-momenta of the Fourier transform of linear perturbations) could be summed only within their radius of convergence. Once we extend the summed-up theories beyond that radius, we often encounter `non-hydrodynamic modes' that are not exactly the same as that of the BDNK theory\footnote{A similar situation arises in the case of the MIS theory as we have presented in section \ref{section MIS}. In the hydrodynamic regime, the stress tensor must be described in a derivative expansion, which turns out to have an infinite number of terms (see equation\eqref{pisum}). Now, in the frequency space ($\omega$), this infinite sum can be performed only within a radius of convergence, which in this case turns out to be $$D\sim |\omega|\leq {1\over \tau_\pi}~.$$
 Introducing new `non-fluid' variables $\pi^{\mu\nu}$ essentially amounts to extending the theory beyond this radius of convergence. Now $\omega =- {i\over\tau_\pi}$ is the new non-hydro mode that emerges in the process of integrating in $\pi^{\mu\nu}$ and this mode is exactly on the radius of convergence of the previous derivative expansion.}.
However, in this section, we shall see that the hydrodynamic modes of the system of equations described in the previous two sections are both exactly the same as that of the BDNK theory at every order in $k$ expansion. This is a consistency test of our claim that our system of equations is indeed equivalent to  BDNK formalism, at least in the hydrodynamic regime.\\

\subsection{Method - 1}

Here, the equivalent system is described by an infinite number of variables and, therefore, an infinite number of equations. For convenience, let us first quote the equations here again.
\begin{align}
 &\partial_\mu T^{\mu\nu}=0~,~~~~~~~~
 T^{\mu\nu}=\hat{\varepsilon}\left[\hat{u}^{\mu}\hat{u}^{\nu}+\frac{1}{3}\hat{\Delta}^{\mu\nu}\right]+\hat{\pi}^{\mu\nu},\nn
 &(1+\tilde\theta\hat D)\hat{\pi}^{\mu\nu}=-2\eta\hat{\sigma}^{\mu\nu}+\rho_1^{\mu\nu},\nn
 &(1+\tilde\chi\hat D)\rho^{\mu\nu}_1=(-2\eta)(-\tilde{\theta})\frac{1}{\hat{T}}\hat{\nabla}^{\langle\mu}\hat{\nabla}^{\rangle\nu} \hat{T}+\rho_2^{\mu\nu},\nn
 &(1+\tilde\theta\hat D)\rho^{\mu\nu}_2=(-2\eta)(-\tilde{\theta})\left(-\frac{\tilde{\chi}}{3}\right)\hat{\nabla}^{\langle\mu}\hat{\nabla}^{\rangle\nu} \hat{\nabla}\cdot\hat{u}+\rho_3^{\mu\nu},\nn
 &(1+\tilde\chi\hat D)\rho^{\mu\nu}_3=(-2\eta)(-\tilde{\theta})^2\left(-\frac{\tilde{\chi}}{3}\right)\frac{1}{\hat{T}}\hat{\nabla}^{\langle\mu}\hat{\nabla}^{\rangle\nu} \hat{\nabla}^2 \hat{T}+\rho_4^{\mu\nu},\nn
 &(1+\tilde\theta\hat D)\rho^{\mu\nu}_4=(-2\eta)(-\tilde{\theta})^2\left(-\frac{\tilde{\chi}}{3}\right)^2\hat{\nabla}^{\langle\mu}\hat{\nabla}^{\rangle\nu}\hat{\nabla}^2\hat{\nabla}\cdot\hat{u}+\cdots\nn
 &\vdots~~~~~~
 \label{MIS-type1-rep}
\end{align}
The `non-fluid' variables are $\pi^{\mu\nu}$ and the infinite sequence of $\rho_n^{\mu\nu}$s, each satisfying a relaxation type of equation.
 
We parameterize the perturbation around static global equilibrium in the following fashion,
\begin{align}
\hat T &= T_0 + \epsilon~ \delta T  ~e^{i T_0(-\omega t + k x)}\nn
\hat u^\mu &= \{1,0,0,0\} +\epsilon~ \{0,\beta _x,\beta_y,0\}  ~e^{i T_0(-\omega t + k x)}\nn
\rho_n^{xx} &=\epsilon~\delta\rho_n^{xx} ~e^{i T_0(-\omega t + k x)}= -2\rho_n^{yy}=-2\rho_n^{zz}~~~~\forall n\nn
\rho_n^{xy} &=\epsilon~ \delta\rho_n^{xy} ~e^{i T_0(-\omega t + k x)}~~~~\forall n~,
\label{MISlinpert2}
\end{align}
all other components of $\rho_n^{\mu\nu}$ vanish for every $n$.

Here, $\epsilon$ is a book-keeping parameter for linearization. Any term quadratic or higher order in $\epsilon $ will be ignored. We have scaled the frequency and the spatial momenta with the equilibrium temperature $T_0$ so that both $\omega$ and $k$ are dimensionless. Similarly, we introduce new dimensionless  parameters of the theory $\tilde\eta_0$, $\tilde \chi_0$ and $\tilde \theta_0$ as follows,
$$\tilde \eta \equiv{ \tilde\eta_0\over T_0},~~~~~\tilde \chi \equiv{ \tilde\chi_0\over T_0},~~~~~\tilde \theta \equiv{ \tilde\theta_0\over T_0}~.$$
If we substitute the perturbations \eqref{MISlinpert2} in equations \eqref{MIS-type1-rep}, we find the dispersion polynomial ${\cal P}(\omega,k)$ whose zeroes will give the modes where the perturbations can have a nontrivial solution.

Now, in this case, it is difficult to express ${\cal P}(\omega,k)$ in a compact form since the equations involve an infinite number of variables. Instead, we shall determine the dispersion polynomial ${\cal P}_N(\omega,k)$ for the same system, truncated at some arbitrary but finite order $n=N$  recursively. The infinite $N$ limit of ${\cal P}_N(\omega,k)$ will give the actual dispersion polynomial of the system. We have,
\be
{\cal P}_N(\omega,k)={\cal P}^\text{shear}(\omega,k)~{\cal P}^\text{sound}_N(\omega,k)~,
\label{MIDdis0}
\ee
where,
 
\begin{align}
{\cal P}^\text{shear}(\omega,k)&= \tilde\eta_0 ~k^2 -i\omega ~(1- i~\tilde\theta_0~\omega)~,\nn
{\cal P}^\text{sound}_N(\omega,k)&=(1 -i~\tilde\chi_0~\omega)^{N\over 2} (1- i~\tilde\theta_0~\omega)^{N\over 2}P_N(\omega,k)~~\text{~When $N$ even,}\nn
{\cal P}^\text{sound}_N(\omega,k)&=(1 -i~\tilde\chi_0~\omega)^{N+1\over 2} (1- i~\tilde\theta_0~\omega)^{N-1\over 2}P_N(\omega,k)~~\text{~When $N$ odd~.}
\label{MIDdis1}
\end{align}
  
Note the  factor $ {\cal P}^\text{shear}(\omega,k)$ is independent of $N$. We could further check that it has the same form as that of the dispersion polynomial in  BDNK theory (see \eqref{BDNK1} and \eqref{BDNK2}) in the shear channel. For $P_N(\omega,k)$ we  have a recursion relation as follows,
 
\begin{equation}\label{recur1}
\begin{split}
P_{2m-1} & = (1 -i~\tilde\chi_0~\omega) P_{2m -2}-i
\left(4 \eta_0\over 3^m\right)~ \tilde\theta_0^m\tilde\chi_0^{m-1}\left( i k\right)^{2(m+1)}~~~~\text{for odd $N=2m-1$},~~m\geq1~,\\
P_{2m} & = (1 -i~\tilde\theta_0~\omega) P_{2m -1}-i
\left(4 \eta_0\over 3^m\right)~ \tilde\theta_0^m\tilde\chi_0^{m}\left( i k\right)^{2(m+1)} (-i\omega)~~~~\text{for even $N=2m$},~~m>0~,\\
P_0&=3 i \omega^2 (1 -i~\tilde\chi_0~\omega) + k^2(i + 4\tilde\eta_0\omega+ \tilde \theta_0\omega)~.
\end{split}
\end{equation}
  
From equations \eqref{MIDdis0}, \eqref{MIDdis1} and \eqref{recur1}, we could see that the degree of the polynomial (and therefore the number of zeroes) in the sound channel increases as we include more and more $\rho_n^{\mu\nu}$s in our system of equations. In other words, with increasing $N$, we keep getting more and more modes. However, it is easy to take $k\rightarrow0$ limit in these recursive equations, and one could see that in the sound channel, there are precisely two modes at $\omega=0$, and all the rests are either at $\bigg[\omega = -{i\over\tilde\chi_0}\bigg]$ or $\bigg[\omega = -{i\over\tilde\theta_0}\bigg]$ similar to the BDNK theory at $k\rightarrow 0$ limit. According to our definitions, the modes with vanishing frequencies at $k\rightarrow 0$ limit are the hydro modes. So, this system of equations does have two hydro modes in the sound channel, as expected from the parent BDNK  theory. Further, by explicit calculation, we can see that these hydrodynamic sound modes match with those of BDNK even at non-zero $k$, if we treat $k$ perturbatively in a power series expansion \footnote{If we truncate the equations at $n=N$, then the frequency of the sound mode matches with that of BDNK upto order {${\cal{O}}\sim (k^{N+3})$}. This we have checked in Mathematica for all $N\leq10$.}. So clearly, the hydro-modes in the equations described in these sections for both the sound and shear channel (in the shear channel, even the non-hydro modes match with BDNK) are the same as those of BDNK, justifying our claim that this system of equations is equivalent to the BDNK systems of equations in the hydrodynamic regime.

\subsection{Method - 2}
\label{method2}
For convenience, let us first quote the system of equations that we would like to analyze,
 \begin{align}
 &\partial_\mu T^{\mu\nu}=0~,~~~~
 T^{\mu\nu}=\hat{\varepsilon}\left[\hat u^{\mu}\hat u^{\nu}+\frac{1}{3}\hat\Delta^{\mu\nu}\right]+\hat{\pi}^{\mu\nu}~,
 \label{BDNKrep1}\\
&\left[(1+\tilde{\theta}\hat{D})(1+\tilde{\chi}\hat{D})-\tilde{\theta}\frac{\tilde{\chi}}{3}~\hat{\nabla}^2\right]
\left\{(1+\tilde{\theta}\hat{D})\hat{\pi}^{\mu\nu}+2\eta\hat{\sigma}^{\mu\nu}\right\}
\nn
&~~~~~~ =~
2\eta\tilde{\theta}\left\{1+(\tilde{\theta}+\tilde{\chi})\hat{D}\right\}\frac{\hat{\nabla}^{\langle\mu}\hat{\nabla}^{\nu\rangle}\hat{T}}{\hat{T}}~.
\label{BDNKrep2}
\end{align}
As before, we parameterize the perturbation around static global equilibrium in the following fashion,
\begin{equation}\label{linpert2}
\begin{split}
\hat T &= T_0 + \epsilon~ \delta T  ~e^{i T_0(-\omega t + k x)}~,\\
\hat u^\mu &= \{1,0,0,0\} +\epsilon~ \{0,\beta _x,\beta_y,0\}  ~e^{i T_0(-\omega t + k x)}~,\\
\pi^{xx} &=\epsilon~\delta\pi^{xx} ~e^{i T_0(-\omega t + k x)}= -2\pi^{yy}=-2\pi^{zz}~,\\
\pi^{xy} &=\epsilon~ \delta\pi^{xy} ~e^{i T_0(-\omega t + k x)}~,\\
\end{split}
\end{equation}
with $\epsilon$ as a book-keeping parameter for linearization. Any term quadratic or higher order in $\epsilon $ will be ignored. Again, the frequency and the spatial momenta are scaled with the equilibrium temperature $T_0$ so that both $\omega$ and $k$ are dimensionless. And also we have introduced new dimensionless  parameters of the theory $\tilde\eta_0$, $\tilde \chi_0$ and $\tilde \theta_0$ as follows,
$$\tilde \eta \equiv{ \tilde\eta_0\over T_0},~~~~~~\tilde \chi \equiv{ \tilde\chi_0\over T_0},~~~~~~\tilde \theta \equiv{ \tilde\theta_0\over T_0}~.$$
Substituting equation \eqref{linpert2} in the system of equations (\eqref{BDNKrep1} and \eqref{BDNKrep2}), we find the following dispersion polynomial,
 
\begin{equation}\label{dispoly2}
\begin{split}
P(\omega, k) = (1 - i \tilde\theta_0\omega)\left[\left(\tilde\chi_0\tilde\theta_0\over 3\right)k^2+ (1 - i \tilde\theta_0\omega)(1 - i \tilde\chi_0\omega) \right] P_\text{BDNK}(\omega,k)~,
\end{split}
\end{equation}
  
where $P_\text{BDNK}(\omega,k)$ is the similar dispersion polynomial computed for the perturbations around the static equilibrium solutions in BDNK systems of equations as given in \eqref{BDNK1} and \eqref{BDNK2} and given by,
 
\begin{equation}
\begin{split}
    P_{\text{BDNK}}=& \left( \tilde\eta_0 k^2 - i \omega (1-i \tilde\theta_0 \omega )\right) \times\\
    &\bigg[ \tilde{\chi}_0\tilde{\theta}_0\omega^4 +i\left(\tilde{\chi}_0+\tilde{\theta}_0\right)\omega^3-\left\{1+\frac{2}{3}\tilde{\chi}_0\left(\tilde{\theta}_0+2\tilde{\eta}_0\right)k^2\right\}\omega^2 \bigg.\\
    \bigg.&-\frac{i}{3}\left(\tilde{\chi}_0+\tilde{\theta}_0+4\tilde{\eta}_0\right)\omega k^2 +\frac{k^2}{3}+\frac{\tilde{\theta}_0}{9}\left(\tilde{\chi}_0-4\tilde{\eta}_0\right)k^4
    \bigg]~.
\end{split}
\end{equation}
  
In other words, the zeroes of $P_\text{BDNK}(\omega,k)$ are the hydro and non-hydro modes of the BDNK theory.

From equation \eqref{dispoly2}, it is clear that all the modes of the BDNK system are already contained in the system of equations \eqref{BDNKrep1} and \eqref{BDNKrep2}. However, they also contain some new modes, which are the zeros of the prefactor,
 
\be
P_\text{extra}(\omega,k)\equiv \left(P(\omega,k)\over P_\text{BDNK}(\omega,k)\right) =(1 - i \tilde\theta_0\omega)\left[\left(\tilde\chi_0\tilde\theta_0\over 3\right)k^2+ (1 - i \tilde\theta_0\omega)(1 - i \tilde\chi_0\omega) \right]~.
\ee
  
Note that all these new modes are of non-hydro type. One could further check that they correspond to the zero modes of the linear PDEs that determine the shift of the velocity and the temperature field ($\delta u^\mu$ and $\delta T$ respectively) under frame transformation (see equation \eqref{shift-soln}).

The existence of such zero modes implies that if we view $\delta u^\mu$ or $\delta T$ as generated from a field redefinition (and not as new `non-fluid' variables), then even after fixing the Landau frame condition, there are still some unfixed residual ambiguities (which exist only at some special form of $\omega(k)$) in the definition of the fluid variables.
On the other hand, if we absorb these shift fields ($\delta u^\mu$ and $\delta T$) into new `nonfluid' variables, the extra zeros of the prefactor  $P_\text{extra}(\omega,k)$ do become the new modes of the theory. In some sense, the residual ambiguities in the field redefinition procedure translate to the
non-uniqueness of the  UV degrees of freedom beyond the hydrodynamic regime.

\section{Conclusion}\label{section Conclude}
In this work, we rewrite the stress tensor of the BDNK hydrodynamic theory in the Landau frame, at least for the part that will contribute to the spectrum of linearized perturbation around static equilibrium. Though the BDNK formalism has a stress tensor with a finite number of derivatives in a generalized frame, it turns out that in the Landau frame, it will have either an infinite number of derivatives or one has to introduce new non-fluid variables. There is no unique way to introduce these non-fluid variables. Here, motivated by the structure of the MIS formalism, we have presented two different ways of doing it, resulting in two completely different-looking sets of equations. However, both the sets have the same hydrodynamics modes as the BDNK theory. But in the process of `integrating in' the non-fluid variables, new non-hydrodynamic modes are generated.

In both methods, we need to do a formal infinite sum over derivatives. We suspect that the convergence issues of these infinite sums, also related to the `non-invertibility' of the zero modes of the linear operator involved in field redefinition, are responsible for these new non-hydrodynamic modes. However, this point needs further investigation. 

More generally, it would be interesting to know if we can identify a part of the spectrum to be invariant under field redefinition and, therefore, truly physical. In this context, the following observation seems useful. In BDNK theory, if we set viscosity ($\eta$) to zero (with nonzero $\chi$ and $\theta$), then via field redefinition, the stress tensor could be made identical to that of an ideal fluid at the linearized level, though in the original `BDNK' frame it will have nontrivial dispersion relation dependent on the values of $\chi$ and $\theta$. This indicates that there might be some partial redundancy in the information contained in the spectrum of a fluid theory. It would be nice to have a more comprehensive understanding of this aspect of the spectrum.
 
Our work has set up a stage for comparison between the BDNK and MIS-type theories. At first glance, they look very different. However, the fluid variables like velocity and temperature used to express the BDNK stress tensor are not the same as the ones used in MIS theory. A comparison is meaningful only if the basic variables of the equations are the same. Once we have done the required transformation, it turns out that though there are differences in the details, the basic structure of nonlocality or `non-fluid' variables is very similar in both theories. The advantage of the Landau frame is that the fluid variables are locally defined in terms of the one-point function of the stress tensor, but in this case, the causal equations turn out to have nonlocal terms or an infinite number of derivatives. Whereas in BDNK theories, the equations are local with a finite number of derivatives, but the fluid variables are related to the one-point function of the stress tensor in a very non-trivial and nonlocal fashion.

However, the BDNK formalism contains more information than what we have just discussed. It says that, there can exist causal relativistic hydrodynamic theories, where the non-localities can be completely absorbed via field redefinitions, thus generating local and causal theories with only a finite number of derivatives. Since the final equation that we have derived for $\pi^{\mu\nu}$ is different from the one in MIS theory, it indicates that one could possibly never absorb the non-localities of the MIS theory in the field-redefinitions.

It would be interesting to extend this analysis to full nonlinear order. Also, it would be very informative to know whether and, if yes, how the story changes as one adds higher-order derivative corrections to the BDNK theory.

\section*{Acknowledgements}
S.B. acknowledges B. Withers for helpful discussions and Trinity College, Cambridge, for hospitality while this manuscript was being prepared. The authors would like to acknowledge their debt to the people of India for their steady and generous support of research in the basic sciences.

\paragraph{Funding information}
S.B., S.M. and S.R. acknowledge the Department of Atomic Energy, India, for the ﬁnancial support.

\begin{appendix}
\numberwithin{equation}{section}

\section{Appendix}

\subsection{Detailed calculations of Method-1}\label{app2}

In this section, we will derive the form of the frame transformations in an infinite-order derivative expansion. To begin with, we'll rewrite the transformations of $T$ and $u^\mu$ under frame redefinitions
\begin{equation*}
    T - \hT = \delta T = \sum_{n=1}^\infty \delta T_n,~~~~~ u^\mu - \hu^\mu = \delta u^\mu = \sum_{n=1}^\infty \hu_n^\mu
\end{equation*}

Substituting this into the expression of the stress-tensor and using the Landau-frame condition, the following expressions are obtained for $\delta T_n$ and $\delta u^\mu_n$.
\begin{equation}
    \begin{split}
        \dtemp_1 &= -\ct \left( \frac{\Hat{D}\hT}{\hT} + \frac{1}{3} \Hat{\nabla} \cdot \hu \right) \\
        \dtemp_{n\ge 2} &= -\ct \left( \frac{\dhat \dtemp_{n-1}}{\hT} + \frac{1}{3} \ghat \cdot \du_{n-1} \right) \\
        \du^\mu_1 &= -\tilt \left( \dhat \hu^\mu + \ghat^\mu \hT \right) \\
        \du^\mu_{n\ge 2} &= -\tilt \left( \dhat \du_{n-1}^\mu + \ghat^\mu \dtemp_{n-1} \right)
    \end{split}
\end{equation}

\subsubsection{Transformation of velocity}
Using the forms given above, we can try to express $\du^\mu_n$ in terms of the lower order $\dtemp$s and $\du^\mu_n$s as,
 
\begin{equation}
    \begin{split}
        \du^\mu_n &= (-\tilt) \left[ \dhat \du^
        \mu_{n-1} + \nabla^
        \mu \frac{\dtemp_{n-1}}{\hT}\right] \\
        &= \left[ (-\tilt)^2 \dhat^2 \delta^
        mu_\nu + \frac{\ct\tilt}{3} \ghat_\nu \ghat^\mu \right] \du^\nu_{n-2} + (-\tilt) (-\tilt-\ct) \dhat \frac{\nabla^\mu \dtemp_{n-2}}{\hT} \\
        &= \left[ (-\tilt)^3 \dhat^3 \delta^\mu_\nu + \frac{\ct \tilt}{3} (-2 \tilt - \ct) \dhat \ghat_\nu \ghat^\mu \right] \du^
        nu_{n-3} + (-\tilt) \left[ (\tilt^2 + \ct \tilt + \ct^2) \dhat^2 + \frac{\ct \tilt}{3} \ghat^2 \right] \ghat^\mu \frac{\dtemp_{n-3}}{\hT} \\
        &= \left[  (-\tilt)^4 \dhat^4 \delta^\mu_\nu + \frac{\ct\tilt}{3} (3\tilt^2 + 2 \ct \tilt + \ct^2) \dhat^2 \ghat_\nu \ghat^\mu + \left(\frac{\ct\tilt}{3}\right)^2 \ghat^2 \ghat_\nu \ghat^\mu  \right] \du^\nu_{n-4} \\
        &+ (-\tilt) \left[  -(\tilt^3 + \ct \tilt^2 + \ct^2 \tilt + \ct^3) \dhat^3 + \frac{\ct\tilt}{3} 2 (-\tilt -\ct) \dhat \ghat^2   \right] \ghat^\mu \frac{\dtemp_{n-4}}{\hT} \\
        &= \left[  (-\tilt)^5 \dhat^5 \delta^\mu_\nu - \frac{\ct\tilt}{3} (4\tilt^3+ 3\tilt^2\ct + 2 \ct^2 \tilt + \ct^3) \dhat^3 \ghat_\nu \ghat^\mu + \left(\frac{\ct\tilt}{3}\right)^2 (-3\tilt - 2 \ct) \dhat \ghat^2 \ghat_\nu \ghat^\mu  \right] \du^\nu_{n-5} \\
        &+ (-\tilt) \left[ (\tilt^4 + \tilt^3 \ct + \tilt^2 \ct^2 + \tilt \ct^3 + \ct^4)\dhat^4 + \frac{\ct\tilt}{3} (3 \tilt^2 + 4 \ct\tilt + 3 \ct^2) \dhat^2 \ghat^2 + \left( \frac{\ct\tilt}{3} \right)^2 (\ghat^2)^2   \right] \ghat^\mu \frac{\dtemp_{n-5}}{\hT}
    \end{split}
\end{equation}
In this way, continuing the sequence, $\du^\mu_n$ can be expressed in terms of $\hT$ and $\hu$ as,
\begin{align}\label{dun}
        \du^\mu_n &= (-\tilt \dhat)^n \hu^\mu
        \notag\\
        &+ (-\tilt) \sum_{m=0}^{n-1} c_m \left( \frac{\ct \tilt}{3} \ghat^2 \right)^m \dhat^{n-1-2m} \frac{\ghat^\mu \hT}{\hT}
        \notag\\
        &+ \sum_{m=0}^{n-2} \frac{\ct\tilt}{3}~ d_m \left( \frac{\ct \tilt}{3} \ghat^2 \right)^m \dhat^{n-2-2m} \ghat^\mu \ghat\cdot \hu~,
    \end{align}
where,
\begin{align}
        c_{mn} &= \frac{1}{(m!)^2}\left(- \frac{\partial}{\partial \tilt} \right)^m \left(-\frac{\partial}{\partial \ct} \right)^m  \sum_{l=0}^{n-1} (-\tilt)^l (-\ct)^{n-1-l}, \\
        d_{mn} &= \frac{1}{(m+1)} \left( -\frac{\partial}{\partial \tilt }\right) c_{mn}~.
\end{align}
The expressions in \eqref{vel1}-\eqref{temp4} can be reproduced from this form in \eqref{dun}.
To find $\du^\mu$, we need to sum over all the $\du^\mu_n$s from $n=1$ to $\infty$ as follows,
\begin{equation}
    \begin{split}\label{threesums}
        \du^\mu = \sum_{n=1}^\infty \du^\mu_n
        = \left(\sum_{n=1}^\infty  (-\tilt \dhat)^n \hu^\mu\right) &+ (-\tilt)\left( \sum_{n=1}^\infty  \sum_{m=0}^{n-1} c_m \left( \frac{\ct \tilt}{3} \ghat^2 \right)^m \dhat^{n-1-2m} \right)\frac{\ghat^\mu \hT}{\hT}  \\
        &+ \left(\sum_{n=1}^\infty\sum_{m=0}^{n-2} \frac{\ct\tilt}{3}~ d_m \left( \frac{\ct \tilt}{3} \ghat^2 \right)^m \dhat^{n-2-2m}\right) \ghat^\mu \ghat\cdot \hu ~.
    \end{split}
\end{equation}
Considering the first summation of \eqref{threesums}, we find that it is an infinite summation of the form
\begin{equation}
    \sum_{n=1}^\infty x^n = x ~\sum_{n=0}^\infty x^n = \frac{x}{1-x}~.
\end{equation}
Hence, from the first summation, we get
\begin{equation}
    \left(\sum_{n=1}^\infty (-\tilt \dhat)^n \right) \hu^\mu = \frac{(-\tilt\dhat)}{(1+\tilt\dhat)} \hu^\mu ~.
\end{equation}
The second summation in \eqref{threesums} is actually a nested summation of three different indices as,
\begin{equation}\label{sum2}
     \sum_{n=1}^\infty  \sum_{m=0}^{n-1}\frac{1}{(m!)^2} \left( \frac{\ct \tilt}{3} \ghat^2 \right)^m \dhat^{n-1-2m} \left(- \frac{\partial}{\partial \tilt} \right)^m \left(-\frac{\partial}{\partial \ct} \right)^m~ \left(  \sum_{l=0}^{n-1} (-\tilt)^l (-\ct)^{n-1-l} \right)~.
\end{equation}
Replacing the index $n$ by $N=n-1$, \eqref{sum2} becomes,
\begin{equation}
    \begin{split}
        \sum_{N=0}^\infty  \sum_{m=0}^{N}\frac{1}{(m!)^2} \left( \frac{\ct \tilt}{3} \ghat^2 \right)^m \dhat^{N-2m} \left(- \frac{\partial}{\partial \tilt} \right)^m \left(-\frac{\partial}{\partial \ct} \right)^m~ \left(  \sum_{l=0}^{N} (-\tilt)^l (-\ct)^{N-l} \right) ~.
    \end{split}
\end{equation}
For values $m>N$, we see that $\left(\frac{\partial}{\partial\tilt}\right)^m$ or $\left(\frac{\partial}{\partial\ct}\right)^m$ acting on the summation over $l$ gives 0 as the highest power of $\tilt$ or $\ct$ in the series is $N$ only. So, we can add an infinite no. of such zeros and extend the summation over $m$ to $\infty$ instead of $N$.
\begin{equation}
    \begin{split}
        \sum_{N=0}^\infty  \sum_{m=0}^{\infty}\frac{1}{(m!\dhat)^2} \left( \frac{\ct \tilt}{3} \ghat^2 \right)^m \dhat^{N} \left(- \frac{\partial}{\partial \tilt} \right)^m \left(-\frac{\partial}{\partial \ct} \right)^m~ \left(  \sum_{l=0}^{N} (-\tilt)^l (-\ct)^{N-l} \right)  
    \end{split}
\end{equation}

The summations over $m$ and $N$ now have independent limits; hence, their order can be interchanged, and we can rewrite the summation as
\begin{equation}
    \begin{split}
          \sum_{m=0}^{\infty} \frac{1}{(m!\dhat)^2} \left( \frac{\ct \tilt}{3} \ghat^2 \right)^m  \left(- \frac{\partial}{\partial \tilt} \right)^m \left(-\frac{\partial}{\partial \ct} \right)^m~ \sum_{N=0}^\infty \dhat^{N} \left(  \sum_{l=0}^{N} (-\tilt)^l (-\ct)^{N-l} \right)  
    \end{split}
\end{equation}

The summations over $N$ and $l$ can then be interchanged using the Cauchy product formula
\begin{equation*}
    \left(\sum_{n=0}^\infty a_n \right) \left(\sum_{m=0}^\infty b_l \right) = \sum_{n=0}^\infty \left( \sum_{l=0}^n a_l b_{n-m} \right) 
\end{equation*}
and \eqref{sum2} can now be expressed as
\begin{equation}
    \begin{split}
        &\sum_{m=0}^{\infty} \frac{1}{(m!\dhat)^2} \left( \frac{\ct \tilt}{3} \ghat^2 \right)^m  \left(- \frac{\partial}{\partial \tilt} \right)^m \left(-\frac{\partial}{\partial \ct} \right)^m~   \left( \sum_{N=0}^\infty (-\ct \dhat)^{N} \right) \left(  \sum_{l=0}^{\infty} (-\tilt\dhat)^l  \right) \\
        = &\sum_{m=0}^{\infty} \frac{1}{(m!\dhat)^2} \left( \frac{\ct \tilt}{3} \ghat^2 \right)^m   \left(- \frac{\partial}{\partial \tilt} \right)^m \left(-\frac{\partial}{\partial \ct} \right)^m~   \left( \frac{1}{(1+\ct \dhat)} \frac{1}{(1+\tilt\dhat)}  \right) \\
        = &\sum_{m=0}^{\infty} \frac{1}{(m!\dhat)^2} \left( \frac{\ct \tilt}{3} \ghat^2 \right)^m \left( \frac{m! \dhat^m}{(1+\ct \dhat)^{m+1}} \frac{m! \dhat^m}{(1+\tilt\dhat)^{m+1}}  \right) \\
        = &\sum_{m=0}^\infty \frac{1}{(1+\ct\dhat)} \frac{1}{(1+\tilt\dhat)} \left( \frac{\frac{\ct \tilt}{3} \ghat^2}{(1+\ct\dhat)(1+\tilt\dhat)} \right)^m \\
        = & \frac{1}{(1+\ct\dhat)} \frac{1}{(1+\tilt\dhat)} \frac{1}{1-\frac{\frac{\ct \tilt}{3} \ghat^2}{(1+\ct\dhat)(1+\tilt\dhat)}} \\
        &= \frac{1}{(1+\ct\dhat)(1+\tilt\dhat)-\frac{\ct \tilt}{3} \ghat^2}
    \end{split}
\end{equation}

Now, let us consider the third summation
\begin{equation}
    \begin{split}
        &\sum_{n=1}^\infty\sum_{m=0}^{n-2} \frac{\ct\tilt}{3}~ \frac{1}{(m!)^2}\frac{1}{(m+1)} \left( \frac{\ct \tilt}{3} \ghat^2 \right)^m \left(-\frac{\partial}{\partial\tilt}\right)^{m+1} \left( -\frac{\partial}{\partial \ct}\right)^m  \dhat^{n-2-2m} \left( \sum_{l=0}^{n-1} (-\tilt)^l (-\ct)^{n-1-l} \right)
    \end{split}
\end{equation}

Here, we see that for $m=n-1$, the no. of $\frac{\partial}{\partial \tilt}$ derivatives becomes more than the highest power of $\tilt$ present in the series over $l$, thus making the term corresponding to $m=n-1$ zero. We can add this zero term, and then our sum becomes
\begin{equation}
    \begin{split}
        &\sum_{n=1}^\infty\sum_{m=0}^{n-1} \frac{\ct\tilt}{3}~ \frac{1}{(m!)^2}\frac{1}{(m+1)} \left( \frac{\ct \tilt}{3} \ghat^2 \right)^m \left(-\frac{\partial}{\partial\tilt}\right)^{m+1} \left( -\frac{\partial}{\partial \ct}\right)^m  \dhat^{n-2-2m} \left( \sum_{l=0}^{n-1} (-\tilt)^l (-\ct)^{n-1-l} \right)
    \end{split}
\end{equation}

Using $N=n-1$ as before,
\begin{equation}
    \begin{split}
        &\sum_{N=0}^\infty\sum_{m=0}^{N} \frac{\ct\tilt}{3}~ \frac{1}{(m!\dhat)^2}\frac{1}{(m+1)} \left( \frac{\ct \tilt}{3} \ghat^2 \right)^m \left(-\frac{\partial}{\partial\tilt}\right)^{m} \left( -\frac{\partial}{\partial \ct}\right)^m  \dhat^{-1} \left(-\frac{\partial}{\partial\tilt}\right) \left( \sum_{l=0}^{N} (-\tilt)^l (-\ct)^{N-l}\dhat^{N} \right)
    \end{split}
\end{equation}

Again, extending the sum over $m$ up to $\infty$ and interchanging summations like the previous case, we get
\begin{equation}
    \begin{split}
        &\sum_{m=0}^{\infty} \frac{\ct\tilt}{3}~ \frac{1}{(m!\dhat)^2}\frac{1}{(m+1)} \left( \frac{\ct \tilt}{3} \ghat^2 \right)^m \left(-\frac{\partial}{\partial\tilt}\right)^{m} \left( -\frac{\partial}{\partial \ct}\right)^m  \dhat^{-1} \left(-\frac{\partial}{\partial\tilt}\right) \left(\sum_{N=0}^\infty \sum_{l=0}^{N} (-\tilt)^l (-\ct)^{N-l}\dhat^{N} \right) \\
        = &\sum_{m=0}^{\infty} \frac{\ct\tilt}{3}~ \frac{1}{(m!\dhat)^2}\frac{1}{(m+1)} \left( \frac{\ct \tilt}{3} \ghat^2 \right)^m \left(-\frac{\partial}{\partial\tilt}\right)^{m} \left( -\frac{\partial}{\partial \ct}\right)^m  \dhat^{-1} \left(-\frac{\partial}{\partial\tilt}\right) \left( \frac{1}{(1+\tilt\dhat)(1+\ct\dhat)} \right)\\
        = &\sum_{m=0}^{\infty} \frac{\ct\tilt}{3}~ \frac{1}{(m!\dhat)^2}\frac{1}{(m+1)} \left( \frac{\ct \tilt}{3} \ghat^2 \right)^m \left(-\frac{\partial}{\partial\tilt}\right)^{m} \left( -\frac{\partial}{\partial \ct}\right)^m  \dhat^{-1}  \left( \frac{\dhat}{(1+\tilt\dhat)^2} \frac{1}{(1+\ct\dhat)} \right)\\
        = &\sum_{m=0}^{\infty} \frac{\ct\tilt}{3}~ \frac{1}{(m!\dhat)^2}\frac{1}{(m+1)} \left( \frac{\ct \tilt}{3} \ghat^2 \right)^m    \left( \frac{(m+1)!\dhat^m}{(1+\tilt\dhat)^{m+2}} \frac{m!\dhat^{m+1}}{(1+\ct\dhat)^m} \right) \\
        = &\frac{\ct\tilt}{3}~\frac{1}{(1+\tilt\dhat)^2}\frac{1}{(1+\ct\dhat)} \sum_{m=0}^{\infty}  \left( \frac{\ct \tilt}{3} \ghat^2 \right)^m    \left( \frac{1}{(1+\tilt\dhat)^m} \frac{1}{(1+\ct\dhat)^m} \right) \\
        = &\frac{\ct\tilt}{3}~\frac{1}{(1+\tilt\dhat)} \frac{1}{(1+\ct\dhat)(1+\tilt\dhat)-\frac{\ct \tilt}{3} \ghat^2}
    \end{split}
\end{equation}

So, putting all these results together, \eqref{threesums} becomes
\begin{equation}
    \begin{split}
        \du^\mu &= \frac{(-\tilt\dhat)}{(1+\tilt\dhat)} \hu^\mu + (-\tilt) \frac{1}{(1+\ct\dhat)(1+\tilt\dhat)-\frac{\ct \tilt}{3} \ghat^2} \frac{\ghat^\mu \hT}{\hT} \\
        &+ \frac{\ct\tilt}{3}~\frac{1}{(1+\tilt\dhat)} \frac{1}{(1+\ct\dhat)(1+\tilt\dhat)-\frac{\ct \tilt}{3} \ghat^2} \ghat^\mu \ghat\cdot\hu \\
        \Rightarrow \du^\mu &= \frac{(-\tilt\dhat)}{(1+\tilt\dhat)} \hu^\mu + \frac{1}{(1+\ct\dhat)(1+\tilt\dhat)-\frac{\ct \tilt}{3} \ghat^2} \left( -\tilt \frac{\ghat^\mu \hT}{\hT} + \frac{\ct\tilt}{3} \frac{1}{(1+\tilt\dhat)} \ghat^\mu \ghat\cdot \hu \right)
    \end{split}
\end{equation}
which we see is identical to the $\du^\mu$ calculated in \eqref{ogv2}.

Also worth noticing is the point that, had we not summed over $m$ in the second and third summations, then $\du^\mu$ would have been left in the form of an infinite series of the form

\begin{equation}
    \begin{split}
        \du^\mu &= \frac{(-\tilt\dhat)}{(1+\tilt\dhat)} \hu^\mu + \frac{1}{(1+\tilt\dhat)(1+\ct\dhat)} \sum_{m=0}^\infty \left( \frac{\frac{\ct\tilt}{3}\ghat^2}{(1+\tilt\dhat)(1+\ct\dhat)} \right)^m \left\{ -\tilt\frac{\ghat^\mu \hT}{\hT} + \frac{\frac{\ct\tilt}{3}}{(1+\tilt\dhat)} \ghat^\mu \ghat\cdot \hu \right\} \\
        u^\mu &= \hu^\mu + \du^\mu = \frac{1}{(1+\tilt\dhat)} \hu^\mu \\
        &+ \frac{1}{(1+\tilt\dhat)(1+\ct\dhat)} \sum_{m=0}^\infty \left( \frac{\frac{\ct\tilt}{3}\ghat^2}{(1+\tilt\dhat)(1+\ct\dhat)} \right)^m \left\{ -\tilt\frac{\ghat^\mu \hT}{\hT} + \frac{\frac{\ct\tilt}{3}}{(1+\tilt\dhat)} \ghat^\mu \ghat\cdot \hu \right\} \\
        \sigma^{\mu\nu} &= \frac{1}{(1+\tilt\dhat)} -2\eta\hat\sigma^{\mu\nu} \\
         & -2\eta \frac{1}{(1+\tilt\dhat)}\frac{1}{(1+\ct\dhat)}  \sum_{m=0}^\infty \left( \frac{\frac{\ct\tilt}{3}\ghat^2}{(1+\tilt\dhat)(1+\ct\dhat)} \right)^m \left\{ -\tilt \frac{\ghat^{\langle\mu}\ghat^{\nu\rangle} \hT}{\hT} + \frac{\frac{\ct\tilt}{3}}{(1+\tilt\dhat)} \ghat^{\langle\mu} \ghat^{\nu\rangle} \ghat\cdot \hu \right\}
    \end{split}
\end{equation}

We can recast this form of $\sigma^{\mu\nu}$ into the form of a relaxation equation given by
\begin{equation}
    \begin{split}
        (1+\tilt\dhat)\sigma^{\mu\nu} &= -2\eta  \left[\hat\sigma^{\mu\nu} 
         + \frac{1}{(1+\ct\dhat)}  \sum_{m=0}^\infty \left( \frac{\frac{\ct\tilt}{3}\ghat^2}{(1+\tilt\dhat)(1+\ct\dhat)} \right)^m \left\{-\tilt \frac{\ghat^{\langle\mu}\ghat^{\nu\rangle} \hT}{\hT} + \frac{\frac{\ct\tilt}{3}}{(1+\tilt\dhat)} \ghat^{\langle\mu} \ghat^{\nu\rangle} \ghat\cdot \hu \right\}\right] \\
         &= -2\eta\hat\sigma^{\mu\nu} + \rho_1^{\langle\mu\nu\rangle}
    \end{split}
\end{equation}

where $\rho_1^{\langle\mu\nu\rangle}$ is given by 
\begin{equation}
    \begin{split}
        \rho_1^{\langle\mu\nu\rangle} &= \left[  
         -2\eta \frac{1}{(1+\ct\dhat)}  \sum_{m=0}^\infty \left( \frac{\frac{\ct\tilt}{3}\ghat^2}{(1+\tilt\dhat)(1+\ct\dhat)} \right)^m \left\{-\tilt \frac{\ghat^{\langle\mu}\ghat^{\nu\rangle} \hT}{\hT} + \frac{\frac{\ct\tilt}{3}}{(1+\tilt\dhat)} \ghat^{\langle\mu} \ghat^{\nu\rangle} \ghat\cdot \hu \right\}\right]
        \end{split}
\end{equation}

It can again be recast into a relaxation equation as
\begin{equation}
    \begin{split}
         \Rightarrow (1+\ct\dhat)\rho_1^{\langle\mu\nu\rangle} &= \left[  
         -2\eta \sum_{m=0}^\infty \left( \frac{\frac{\ct\tilt}{3}\ghat^2}{(1+\tilt\dhat)(1+\ct\dhat)} \right)^m \left\{ -\tilt \frac{\ghat^{\langle\mu}\ghat^{\nu\rangle} \hT}{\hT} + \frac{\frac{\ct\tilt}{3}}{(1+\tilt\dhat)} \ghat^{\langle\mu} \ghat^{\nu\rangle} \ghat\cdot \hu \right\}\right] \\
         &= -2\eta (-\tilt) \frac{\ghat^{\langle\mu}\ghat^{\nu\rangle} \hT}{\hT} + \rho_2^{\langle\mu\nu\rangle}
    \end{split}
\end{equation}

with $\rho_2^{\langle\mu\nu\rangle}$ defined and associated with another relaxation equation as
\begin{equation}
    \begin{split}
        &\rho_2^{\langle\mu\nu\rangle} =  \frac{-2\eta\frac{\ct\tilt}{3}}{(1+\tilt\dhat)} \ghat^{\langle\mu} \ghat^{\nu\rangle} \ghat\cdot \hu - 2\eta \sum_{m=1}^\infty \left( \frac{\frac{\ct\tilt}{3}\ghat^2}{(1+\tilt\dhat)(1+\ct\dhat)} \right)^m \left\{ -\tilt \frac{\ghat^{\langle\mu}\ghat^{\nu\rangle} \hT}{\hT} + \frac{\frac{\ct\tilt}{3}}{(1+\tilt\dhat)} \ghat^{\langle\mu} \ghat^{\nu\rangle} \ghat\cdot \hu \right\} \\
        &(1+\tilt\dhat) \rho_2^{\langle\mu\nu\rangle} = -2\eta \frac{\ct\tilt}{3} \ghat^{\langle\mu} \ghat^{\nu\rangle} \ghat\cdot \hu + \rho_3^{\langle\mu\nu\rangle}
    \end{split}
\end{equation}

where again $\rho_3^{\langle\mu\nu\rangle}$ contains the infinite series. In this way, the sequence would continue, and any general term would be given by (for $n\ge0$)
\begin{equation}
    \begin{split}
        &(1+\ct \dhat) \rho_{2n+1}^{\langle\mu\nu\rangle} = (-2\eta)(-\tilt) \left( \frac{\ct\tilt}{3} \ghat^2 \right) ^n~ \frac{\ghat^{\langle\mu}\ghat^{\nu\rangle} \hT}{\hT} + \rho_{2n+2}^{\langle\mu\nu\rangle} \\
        &(1+ \tilt\dhat) \rho_{2n+2}^{\langle\mu\nu\rangle} = (-2\eta)\frac{\ct\tilt}{3} \left( \frac{\ct\tilt}{3} \ghat^2 \right) ^n~  \ghat^{\langle\mu} \ghat^{\nu\rangle} \ghat\cdot \hu + \rho_{2n+3}^{\langle\mu\nu\rangle} \\
        &\rho_{2n+1}^{\langle\mu\nu\rangle} =  \frac{-2\eta}{(1+\ct\dhat)} \sum_{m=n}^\infty \left( \frac{\frac{\ct\tilt}{3}\ghat^2}{(1+\tilt\dhat)(1+\ct\dhat)} \right)^m \left\{ -\tilt \frac{\ghat^{\langle\mu}\ghat^{\nu\rangle} \hT}{\hT} + \frac{\frac{\ct\tilt}{3}}{(1+\tilt\dhat)} \ghat^{\langle\mu} \ghat^{\nu\rangle} \ghat\cdot \hu \right\} \\
        &\rho_{2n+2}^{\langle\mu\nu\rangle} = \frac{-2\eta}{(1+\tilt\dhat)}  \frac{\ct\tilt}{3} \left( \frac{\ct\tilt}{3}\ghat^2\right)^n~ \ghat^{\langle\mu} \ghat^{\nu\rangle} \ghat\cdot \hu \\
        &+  \sum_{m=n+1}^\infty \left( \frac{\frac{\ct\tilt}{3}\ghat^2}{(1+\tilt\dhat)(1+\ct\dhat)} \right)^m \left\{ -\tilt \frac{\ghat^{\langle\mu}\ghat^{\nu\rangle} \hT}{\hT} + \frac{\frac{\ct\tilt}{3}}{(1+\tilt\dhat)} \ghat^{\langle\mu} \ghat^{\nu\rangle} \ghat\cdot \hu \right\}
    \end{split}
\end{equation}
These are the general forms of the $\rho_n^{\langle\mu\nu\rangle}$s given in \eqref{MIS-type1}.

\subsubsection{Transformation of temperature}

As it was done in the previous subsection, the expression for $\frac{\dtemp_n}{\hT}$ can be written as,
 
\begin{equation}
    \begin{split}
        \frac{\dtemp_n}{\hT} &=  \frac{(-\ct \dhat)^n \hT}{\hT} + (-\ct) \sum_{m=0}^{n-1} c_{mn} \dhat^{n-1-2m} \left( \frac{\ct\tilt}{3} \ghat^2 \right)^m \left(\frac{\ghat\cdot \hu}{3}\right) \\
        &+ \frac{\ct \tilt}{3} \sum_{m=0}^{n-2} f_{mn} \dhat^{n-2-2m} \left( \frac{\ct\tilt}{3} \ghat^2 \right)^m \frac{\ghat^2\hT}{\hT} 
    \end{split}
\end{equation}
where $c_{mn}$ is defined the same way as in $\du^\mu_n$ and $f_{mn}$ is defined in terms of $c_{mn}$ as
\begin{equation}
    f_{mn} = \frac{1}{m+1} \left( - \frac{\partial}{\partial\ct} \right) c_{mn}
\end{equation}

Similar to the case of $\du^\mu$, we again take an infinite summation over $n$ to obtain $\frac{\dtemp}{\hT}$ as
\begin{equation}
    \begin{split}
        \frac{\dtemp}{\hT} = \sum_{n=1}^\infty \frac{\dtemp_n}{\hT} = &-\frac{\ct}{(1+\ct \dhat)} \frac{\dhat\hT}{\hT} + (-\ct) \frac{1}{(1+\ct\dhat)(1+\tilt\dhat)-\frac{\ct \tilt}{3} \ghat^2} \left( \frac{\ghat\cdot\hu}{3} \right) \\
        &+ \frac{\ct\tilt}{3} \frac{1}{(1+\ct \dhat)} \frac{1}{(1+\ct\dhat)(1+\tilt\dhat)-\frac{\ct \tilt}{3} \ghat^2} \frac{\ghat^2 \hT}{\hT}
    \end{split}
\end{equation}

and from there, obtain the same $T = \hT + \dtemp$ as in \eqref{ogt2}
\begin{equation}
    \begin{split}
        T &= \frac{1}{(1+\ct\dhat)} \hT -\hT\ct  \frac{1}{(1+\ct\dhat)(1+\tilt\dhat)-\frac{\ct \tilt}{3} \ghat^2} \left( \frac{\ghat\cdot\hu}{3} \right) +  \frac{1}{(1+\ct \dhat)} \frac{1}{(1+\ct\dhat)(1+\tilt\dhat)-\frac{\ct \tilt}{3} \ghat^2} \frac{\ct\tilt}{3}\ghat^2 \hT \\
        &= \frac{1}{(1+\Tilde{\theta}D) (1+ \Tilde{\chi} D) - \frac{\Tilde{\chi} \Tilde{\theta}}{3} \nabla^2 } \left[ (1+\tilt\dhat) \hT - \hT \ct \frac{\ghat\cdot\hu}{3} \right]
    \end{split}
\end{equation}
  
\end{appendix}

\end{document}